# A Comparative Study of Image Denoising Algorithms

Muhammad Umair Danish

COMSATS Institute of Information Technology Islamabad 44000, Pakistan

Emails: {udanish508}@gmail.com

**Abstract:** With the recent advancements in the field of information industry, critical data in the form of digital images is best understood by the human brain. Therefore, digital images play a significant part and backbone role in many areas such as image processing, vision computing, robotics, and bio-medical. Such use of digital images is practically implementable in various real-time scenarios like biological sciences, medicine, gaming technology, computer information and communication technology, data and statistical science, radiological sciences and medical imaging technology, and medical lab technology. However, when any digital image is sent electronically or captured via camera, it is likely to get corrupted or degraded by the available of degradation factors. To eradicate this problem, several image denoising algorithms have been proposed in the literature focusing on robust, low-cost and fast techniques to improve output performance. Consequently, in this research project, an earnest effort has been made to study various image denoising algorithms. A specific focus is given to the start-of-the-art techniques namely: NL-means, K-SVD, and BM3D. The standard images, natural images, texture images, synthetic images, and images from other datasets have been tested via these algorithms, and a detailed set of convincing results have been provided for efficient comparison.

## I. Introduction

Latest accelerations toward technological advancements in image processing based Internet-of-Things (IoT) offer great deal of support to the research community dedicated to develop extremely convenient and elegant systems in terms of design, computational cost, and practical implementation. Such efforts have allowed the electronics industry to assemble wireless devices with tiny structure, economical value, and having the ability to efficiently use the available power resources.

However, these electronic gadgets are always accompanied by various algorithms deployed within them. For example, an electronically developed camera needs novel image and video processing algorithms in order to filter out the unwanted processes for improving the performance. This, on the other hand, is always severely disturbed by the additional computational burden added as a side-effect of information processing algorithms.

In this connection, the recently completed researches have focused on a desired trade-off between processing output performance and that of a computational overhead. Currently, one of the trending research areas in this regard is that of analyzing and combatting noisy components in image signals. Since many of the signals travel through wireless media, there is always an inevitable presence of noise which ultimately corrupts these data signals.

To outperform the unwanted noise, many algorithms have been introduced. Consequently, in this thesis, we study various state-of-the-art image denoising algorithms and analyze a number of image denoising results via extensive simulations using MATLAB®. In the following sections, we present motivation and objectives of this project. This is then followed by a hands-on background information about the relevant topics.

a. Motivation

The inspiration for outlining this research comes from recent advancements in the field of image processing algorithms deployed in IoT-based networks. This study lays out detailed study of digital images and analysis of different image denoising algorithms that play a vital role in research and engineering technology.

b. Objectives

The main objectives of our research study are as under:
- To briefly study different types of noises that corrupts images when sent wirelessly
- To study various image denoising techniques
- To implement image denoising algorithms for restoration
- To computer PSNR and SSIM of recovered images
- To make comparative analysis of various noise denoising algorithms

c. Aim and Basic Idea

Generally, the image data transfer is about visual information transmitted in the form of digital images that is more interactive and useful but which need to remove all the noise and degradation from the images. The received image needs processing before it can be used in practical applications such as video recording. This is basic idea of study of image restoration and denoising techniques.

d. Information Goal

Visual or information in the form of digital images is becoming a major method of communication in this modern age and it is being used in different fields of engineering, medical sciences, earth sciences, and even in core area of geographical information systems. High quality images become noisy after transmission using a wireless channel since the channel is equipped with naturally present noise. The received image needs processing before it can be used in applications. Image denoising involves the manipulation of the image data to produce a visually decent and high-quality image that resembles the original scenery.

e. Methodology Introduction

As a critical part of this Study, we have selected various state-of-the-art image denoising algorithms. Our contribution of the thesis is mainly enriched via following key working steps:
- Study and analysis of various image denoising algorithms
- A vast range of simulations carried out in MATLAB$^®$
- Computation of comparison metrics and analysis of results

f. Applications of Research

This research will be applied on images used in daily life application used by information and technology industry such as computer and IT, geographic information systems, medicine and biological sciences. The graphical information is the furthermost significant type of information perceived, processed and interpreted by the human brain. In the field of education, images became a key and compulsory block of instructional processes.

In large-scale enterprises development systems, digital image plays a vital role in understating and analysis of the key data. Images provides facility of graphical reports which is mostly best processed by man. In the biological sciences, images provide support to understand different aspects of micro-level analytics. Images also provide processes in the form of different pictorial flow charts which create easy level of understanding. This is shown in Fig. 1 with the help of examples.

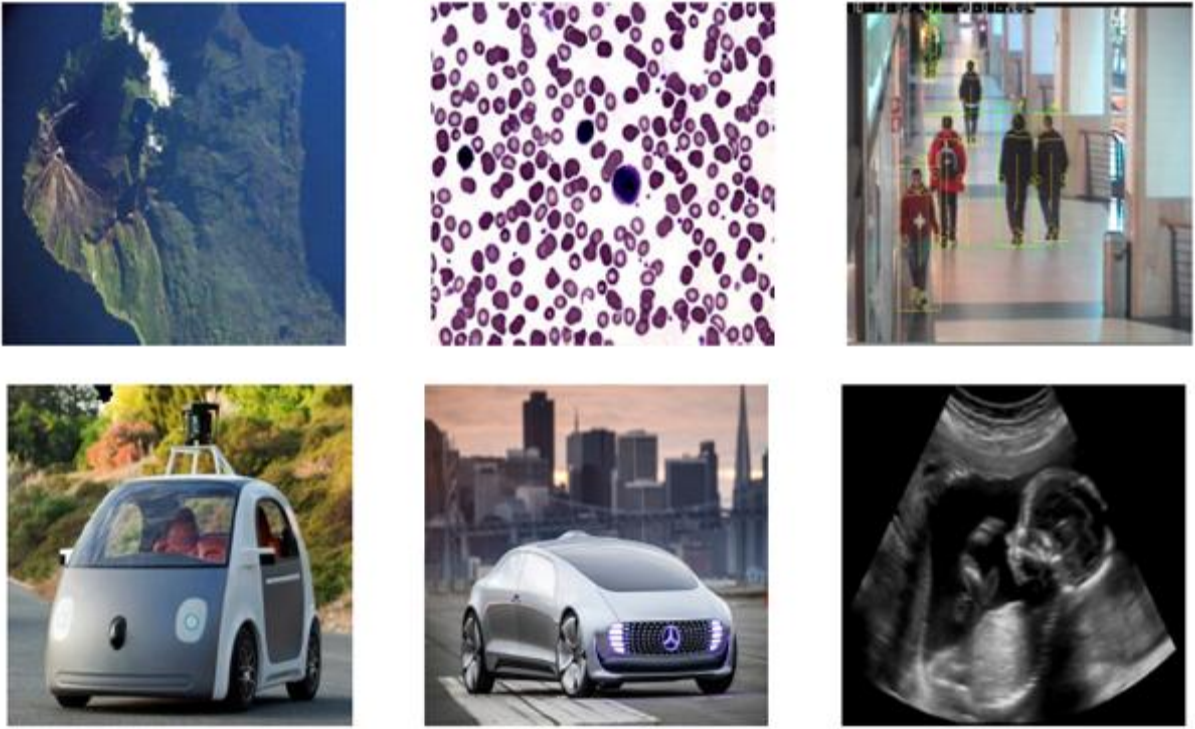

Figure 1: Applications of image processing algorithms in satellite communication, microscopic science, computer vision, bio-medical sciences, and future of transportation. [Courtesy: M. Behzad, Compressed Sensing Based Image Denoising: Novel Patch-Based Collaborative Algorithms, M.S. Electrical Engineering Dissertation, King Fahd University of Petroleum & Minerals – Online Source: http://muzammilbehzad.com]

g. Scope and Limitations

The scope of this project relates it towards the utilization of modern technologies by implementing the denoising techniques on various range of fields that span from engineering, medicine, and many others. Our research project will focus on the comparative analysis of algorithms with respect to overall better output performance. However, the study has majorly focused on the key following items:
1. The proposed work aims to develop a uniform programming code in MATLAB® for implementation and comparison of different denoising algorithms.
2. Three major state-of-the-art algorithms will be studied and compared.
3. The experimentation will be carried out on several types and sizes of images.
4. The behavior of each algorithm against each types and size of image will be presented.
5. A number of comparison graphs will be used to evaluate the results.

Our work has following limitations:
1. The noisy images cannot be fully denoised.
2. The MATLAB® simulation takes lot of time.
3. A strong processing PC is required for simulations.

h. Paper Outline

In this research work, image denoising algorithms recovered images from noisy images. Images get noisy due to several reasons, such as, using a defected camera, sending from one to another device, undesired lightening scenarios while capturing footage. Noise is the corrupt signal that affects the original signal and changes pixel from original position. This is the reason why image gets noisy or degraded. Noise can disturb quality of digital or binary images. There are numerous potential sources of noise. The main reason of noisy signals are low quality camera instruments, defect in operation while sending to another storage medium, and natural lights which affects the quality of image acquiring machine.

The aim of image denoising processing is to improve the possible information for human interpretation. It is also processing image for storage transmission and representation for autonomous machine perception. Digital images are often corrupted by impulse noise in transmission error. The different types of noise are Additive White Gaussian Noise (AWGN), impulse noise, etc. Different types of noise corrupt an image during the process of acquisition, transmission, reception, and storage and retrieval. All these factors combined and made an image noisy. Consequently, after that image needs to be restored before it is used. To complete the purpose denoising techniques are used. In this regard, a lot of research papers have written on the restoration of images corrupted by noise where image deploring and denoising are the two sub areas of image restoration.

i. What are Images?

Images are graphical or visual representation of substance, scenario, process or something. Images are used for better understanding. Image as a term used differently in various fields of knowledge. An image is a picture that has been created, taken by camera or copied from somewhere and stored in electronic form in storage device. An Image is two-dimensional photos have same appearance to the object. Images are stored in many formats, in which some are given below.

1. Grayscale Image

In simple words, it is said "black and white" image; in which value of pixel is 8 bits, or it is binary image having only colors, but these colors also include all shades of gray. In the computing, grayscale image can be calculated through rational numbers whereas image pixels are quantized to store unsigned integers.

Grayscale images are distinct from one-bit bi-tonal black-and-white images, which in the context of computer imaging are images with only two colors, black and white. Grayscale images have many shades of gray having intensity values from 0-255 as shown in Fig. 2.

Grayscale images can be the result of measuring the intensity of light at each pixel according to a particular weighted combination of frequencies or wavelengths, and in such cases they are monochromatic proper when only a single frequency in practice, a narrow band of frequencies is captured. The frequencies can in principle be from anywhere in the electromagnetic spectrum gray scale image is an image that has a defined gray scale color space, which maps the stored numeric sample values to the achromatic channel of a standard color space, which itself is based on measured properties of human vision.

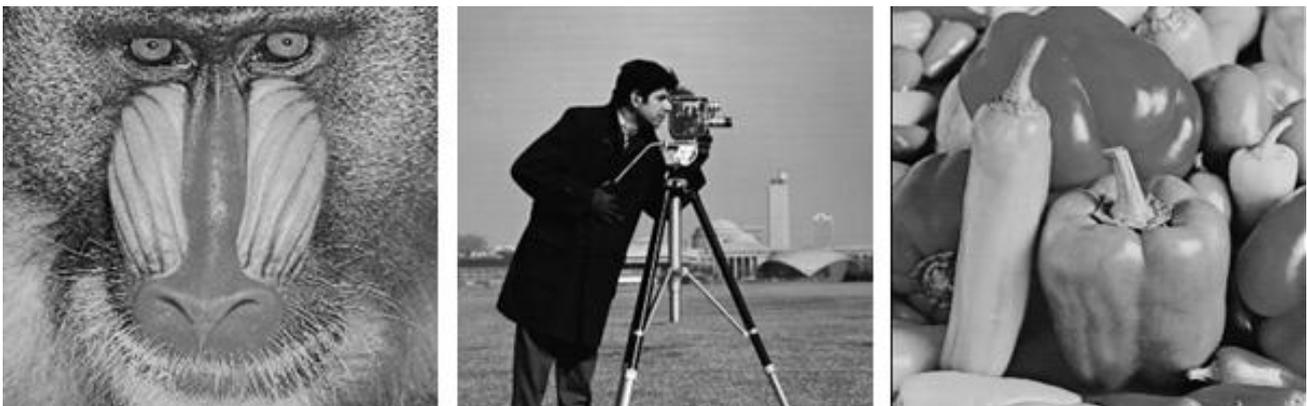

Figure 2: Some examples of standard grayscale images used in image processing having intensity values from 0 – 255
[Left-to-Right: *Mandrill*, *Cameraman*, and *Peppers*]

2. Color Images

A color image is a digital image with color information at different pixels. It is visually better than grayscale image. It is also termed as RGB (red, green, blue) image, and is stored as three-dimensional array where each dimension refers to the R, G, or B channel. This is shown in Fig. 3 with the help of RGB channels.

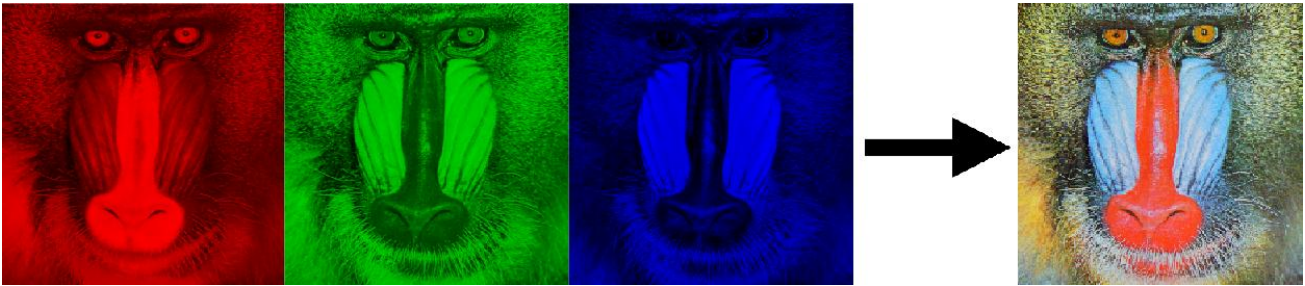
Figure 3: Individual red, green, and blue (RGB) channels to form a colored *Mandrill* Image

j.  Image Degradation

Noise is undesirable signal that affects the original image and degrades the graphic value of image. The key sources of noise in digital images are defective instrument that are used in the process, problems with data acquisition system, interfering normal phenomena, and issues while transmission of images.

The expected image requirements handling before it can be used in applications. Image denoising involves the use of the image data to produce a visually high-quality. The aim of image denoising processing is to improve the possible information for human interpretation. It is also to pre-process image for storage and representation for autonomous machine perception.

k.  Kinds of Noise

1.  Gaussian Noise

"Gaussian noise is disseminated over the signal. This is means that each pixel in the noisy image is the total of the true pixel value and an arbitrary Gaussian disseminated noise value. As the name specifies, this type of noise has a Gaussian distribution, which has a bell-shaped probability distribution."

2.  Salt and Pepper Noise

"Salt and Pepper is the second major type of noise, an urge type of noise and is also stated to as power spikes. It is may be caused due to faults in transmission, image taken by camera is the reason behind all. This is also caused due to errors in data transmission from different instrument. It takes only two possible values, one is a and two is b. The probability is less than 0.1. The degraded pixels are set instead to the less or to the more value, giving the image a "salt and pepper" appearance. Unpretentious pixels endure unchanged. The salt and pepper noise is produced by malfunctioning of pixel basics.

3.  Speckle Noise

"This is third type of noise occurs in almost all coherent imaging systems for example laser, acoustics and SAR Synthetic Aperture Radar imagery. The source of this noise is ascribed to random interference between the coherent returns."

Speckle noise in conventional radar results from random fluctuations in the return signal from an object that is no bigger than a single image-processing element. It increases the mean grey level of a local area.

4.  Brownian Noise

Brownian noise originates in of fractal or 1/f. The mathematical model for 1/f noise is Brownian. Brownian Noise is used often after taking images via camera. Brownian noise is case of 1 out of f is ''noise''. It is taken by assimilating "white noise". The graphic representation of the sound signal mimics a Brownian pattern. Its spectral density is inversely proportional to $f^2$, meaning it has more energy at lower frequencies, even more so than pink noise.

l.  Image Formats

1.  JPEG

JPEG is read as Joint Photographic Expert Group, which is used most commonly rather than any other format. JPEG has 16 Million possible colors which produced using 8 bits for each. JPEG also don't support extra text which is disadvantage.

"A very important implementation of a JPEG codec is the free programming library libjpeg of the Independent JPE Group. It was first published and was key for the success of the standard. This library or a direct derivative of it is used in countless applications."

'The JPEG compression algorithm is at its best on photographs and paintings of realistic scenes with smooth variations of tone and color. "For web usage, where reducing the amount of data used for an image is important for responsive presentation, JPEG's compression benefits make JPEG popular. JPEG is also the most common format saved by digital cameras.

However, JPEG is not being as well suited for line drawings and other textual or iconic graphics, where the sharp contrasts between adjacent pixels can cause noticeable artifacts. Such images are better saved in a lossless graphics format such as TIFF, GIF, PNG, or a raw image format." The JPEG standard includes a lossless coding mode, but that mode is not supported in most products

2. PNG

Portable Network Graphics file format is commonly used for websites or processes where images need to be transfer via network. It provides several improvements.

Pixels in PNG are many numbers that may be directories of sample data in the palette or the sample data itself. The palette is a isolated table delimited. Sample data for a single pixel consists of a tuple of between one and four numbers. Whether the pixel data signifies palette indices or clear instance values, the numbers are denoted to as channels and every number in the image is encoded with an identical format.

The legalized formats encode each number as an unnamed integral worth using a fixed number of bits, raised to in the PNG requirement as the bit depth. Notice that this is not the same as color depth, which is recurrently used to refer to the total number of bits in each pixel, not each channel. The permitted bit depths are undersized in the table along with the total number of bits used for each pixel.

3. TIFF

The TIFF is read as for Tagged Image File format is a flexible that saves 8 bits. TIFF files are commonly used in computer publishing, sending the image via fax, 3 Dimensional applications, and medical imaging's. The labelled structure was designed to be easily extendible, and many vendors have introduced exclusive special-purpose. Adobe technical notes have been published with minor extensions to the format, and several specifications have been based on TIFF 6.0, TIFF readers must be prepared for multiple/multi-page images (sub files) per TIFF file.

"Although they are not required to actually do anything with images after the first one.

here may be more than one Image File Directory (IFD) in a TIFF file. Each IFD defines a sub file. One use of sub files is to describe related images, such as the pages of a facsimile document. A Baseline TIFF reader is not required to read any IFD beyond the first one."

4. GIF

GIF stands for Graphics Interchange Format supports both animated and static images. "The format supports up to 8 bits per pixel for each image, allowing a single image to reference its own palette of up to 256 different colors chosen from the 24-bit RGB color space. It also supports animations and allows a separate palette of up to 256 colors for each frame. These palette limitations make GIF less suitable for reproducing color photographs and other images with color gradients, but it is well-suited for simpler images such as graphics or logos with solid areas of color.

GIF images are compressed using the lossless data compression technique to reduce the file size without degrading the visual quality.

This compression technique was patented. Controversy over the licensing agreement between the software patent holder, Unisys, and CompuServe spurred the development of the Portable Network Graphics (PNG) standard. All the relevant patents had expired."

m. PSNR and SSIM

1. PSNR

"Peak signal-to-noise ratio, name given is PSNR, is used in Image Processing while measuring Image quality. It is used to calculate different PNSR of images, for the ratio amongst the maximum probable power of a signal and the power of corrupting noise that affects the dependability of its representation. If PSNR/ Value will higher, Image will be higher."

2. SSIM

"The Structural Similarity Index (SSIM) is a perceptual metric that quantifies image quality degradation caused by processing such as data compression or by losses in data transmission. It is a full reference metric that requires *two* images from the same image capture a reference image and a processed image. The processed image is typically compressed. It may, for example, be obtained by saving a reference image as a JPEG at any quality level) then reading it back in. SSIM is best known in the video industry, but has strong applications for still photography."

n. Summary

Digital Images are becoming very important in all fields of life. These images play key role in Business and Marketing, Electronics, Engineering, Biological and Medical Sciences. In this section, we have enclosed comprehensive introduction of images, image noise and degradation, types of images, types of noises, etc. The aim of an image denoising processing is to improve the possible information for human interpretation. It is also processing image for storage and representation for autonomous machine perception. Digital images are often corrupted by AWGN noise in transmission. Different types of noise corrupt an image during the process of acquisition, transmission, reception, and storage and retrieval.

Consequently, after that image needs to be restored before it is used. For this, the denoising techniques are used. A lot of research have been carried out on the restoration of images corrupted by noise where image deploring and image de noising are the two sub areas of image restoration. In this thesis work, we are investigating image denoising algorithms applied on various images that are corrupted by AWGN.

## II. LITERATURE REVIEW

This section delivers the detailed contextual of the exploration conducted and discussions are made on all the techniques used in this project. The section will be further divided into sub-sections that provides with the brief introduction of the organization. The main Information is to be depicted here.

Image processing is an area that continues to evolve and evolve with increasing speed. It is a fascinating and exciting field with many applications ranging from the entertainment industry to the space program. One of the most interesting aspects of the informal revolution is the excitement of carrying and receiving complex data outside the plain text. Visual notes in the form of digital images have become an important communication tool for the 21st century. Image processing is the type of signal processing that is an image input, including an image or a video, and the image processing output may be an image or a set of associated properties or parameters. The subsequent sections provide a description of famously adopted techniques in the field of denoising via various image and signal processing techniques.

a. Background

This comparative study looks at the problem of images while transferring ultimately corrupted by noise. The key concerned factors in related research work carried out are:
- There are many algorithms available for denoising images. None has their distinct set of specification to understand a complete framework.
- While denoising images, it is not clear whether algorithms work for all types of images or they are merely proposed for a favorable dataset of images.

Hence, to study this issue, we propose a comparative study of image denoising algorithms, with a soul purpose of removing noise. The main advantages of such an algorithm should be that it should greatly simplify the ordering process for all types of images.

b. Related Work

(S. Suresh and S. Lal-2017) In this research, we recommend a two-dimensional search algorithm (2D-CSAWF) to remove noise from satellite

images contaminated by Gaussian noise. To our knowledge, research based on adaptive two-dimensional Wiener filtering based on meta-heuristic algorithms has not been found in the literature. Comparisons are performed using the latest advanced 2D adaptive noise filter algorithm to analyze the performance and computational efficiency of the proposed algorithm. We have also included a comparison with the recent adaptive metaheuristic algorithms used to suppress noise from satellite images and provide a fair comparison. All algorithms are tested on a set of satellite image data to exclude noise from the damaged image with three different levels of Gaussian noise variation. Experimental results show that the new 2D-CSAWF algorithm proposed surpasses that of others in quantitative and qualitative terms. [1]

(S. Xu, Y. Zhou, H. Xiang and S. Li, 2014) In this article, it is particularly effective to make the algorithm non-local (NLM) because the remote sensing image contains repetitive image patches. Block wise NLM (BNLM) improves the lack of NLM time complexity, but there are still problems with contour blur and loss of detail. In this research, we suggest an NLM algorithm (NLMPG) based on clustering patches. This algorithm tracks BNLM when evaluating the value of a patch based on its similarity to other patches in the image, but only the most similar patch number is selected; it helps to eliminate non-specific information. relevant. of the filter as a whole. NLMPG customizes the filter constant values of each central patch based on the variance proportion of the image patch and provides better performance. Investigational results verify that the suggested NLMPG algorithm is effective for maintaining structure and edge retention and achieving cutting-edge noise reduction performance with respect to quantitative criteria and subjective visual quality. [2]

(C. Aguerrebere, A. Almansa, J. Delon, Y. Gousseau and P. Musé, 2017). In recent years, impressive noise suppression results are obtained using a Bayesian approach including a Gaussian model for image correction. This performance improvement is due to the use of templates in patches. Unfortunately, this approach is particularly instable to most inverse problems, in addition to eliminating noise. In this research, we suggest to use high priority image archetypal to stabilize the evaluation procedure. The proposed recovery scheme has two main advantages: First, it is suitable for diagonal degradation matrices, especially with no data problems (eg, scale). Secondly, we can handle signal dependent noise model, especially noise model suitable for digital camera. Therefore, this method is particularly suitable for calculation photographs. To illustrate this point, we propose an application for a high dynamic image from a single image made with a modified sensor showing the effectiveness of the proposed scheme. [3]

(L. Jia *et al*, 2017) This is the document I read in the K-SVD-based classical representation redundancy algorithm that only drives the dictionary with a fixed atom size for the entire image, which is limited in the precise description of the image. To overcome this deficiency, this paper presents an effective algorithm for deleting images with improved dictionaries. Firstly, on the basis of geometrical and photometric similarities, image patches are grouped into different groups. Secondly, these groups are classified in the flat category, the texture category and the edge category. In different categories, the size of the atoms in the dictionary is designed differently. Therefore, the dictionary of each group is driven with the size of the atom determined by the category to which the group belongs and by the noisy level. Finally, the noise elimination method is presented using a scattered representation in grouped dictionaries constructed with adaptive atomic dimensions. Experimental results show that the proposed method allows to obtain a better noise reduction performance compared to noise reduction algorithms, in particular in the preservation of the image structure. [4]

(A. Karami and L. Tafakori, 2017) In most image dispensation processes, it is important to reduce the noise level. The study aims to introduce effective methods for this purpose based on the general distribution of Cauchy (GC). As a result, some characteristics of the GC distribution are considered. In particular, the GC distribution function is obtained by using the particular theory of positive density and using the density of GC unsystematic variables as a gathering of the sophistication function of the two Linnaeus variables rather than the general symmetry. In addition, the GC distribution is well-thought-out a filter, and in the suggested technique of image noise attenuation, the most favorable considerations of the GC filter are determined by the optimization of the particles of the swarm. The proposed method is used for different types of sound images and the results are compared to the four main complaint algorithms. Investigational outcomes substantiate that their techniques can diminish the impacts of noise. [5]

(Y. Chen, Y. Guo, Y. Wang, D. Wang, C. Peng and G. He, 2017) Hyperspectral imaging removal (HSI) is a challenge not only the difficulty of preserving spectral and spatial structures at the same time, but also the need to eliminate various noises, often cosmopolitan composed. In this research, we proposed a convex array pattern approximation of low range (NonLRMA) and the communicating HSI denoising technique to reformulate the problem approach using a non-convex regulator instead of the standard nuclear energy functions from original interval function to regularized dispersion. NonLRMA aims to split HSI gradient, represented as a matrix, a short-range constituent and a rare term with an additional vigorous and less partial interpretation. We have also developed an iterative algorithm constructed on the updated Lagrange multiplier technique and descend the resulting impassable clarification of sub-problems that benefit from the special non-convex replacement feature. We demonstrate that our iterative optimization converges easily. HSI extensive replicated and authentic experiments designate that our technique can not only suppress band-serious noise and a bit noisy, but also support huge images and small-measure details. Comparison with HSI removing modern LRMA-based approaches showed our superior noise performance. [6]

(J. M. Mejia, H. J. Ochoa, O. O. Vergara, B. Mederos and v. g. cruz, 2017) In this paper, the author presents an algorithm for eliminating noise from insignificant animal positron emission images. The suggested algorithm associations the transformation of multiple resolutions with a reliable filtering of the areas. The image is processed in an area of the non-intersecting contour, using the conversion capabilities to capture the geometric information of important structures, such as minor damage and ribs between the fabrics. Furthermore, in the field of transformation, we have proposed to use virtually stable potentials to reduce noise in areas without borders, this is done by evaluating the border map and the set of image areas. Finally, the reverse cycle transformation is used to produce an image with disturbing effects. Quality tests with the NEMA NU4 2008 puppet indicate that the suggested method decreases noise in the image, while the average is maintained in each region. Comparison with other methods, using contrast analysis in fictitious trauma, shows the superiority of our approach to denaturing and preserving small structures, such as lesions. [7]

(H. He, W. J. Lee, D. Luo and Y. Cao, 2017) In this article, the author has discussed that the methods of infrared detection of faults in the supremacy gridiron have fascinated a lot of consideration in recent years. Because the infrared image of the insulating line has a higher-level noise and a lower divergence, it will imitate the accurateness of the determination of null paddings. In this research, we propose a technique based on general Gaussian bursts and a thoroughgoing estimate of the subsequent possibility of eliminating the noise of electromagnetic insulating images. Because of the high highest and elongated tail characteristics of infrared image wave measurements, the Generalized Gaussian Distribution (GGD) is used as a possibility dissemination function. The concentrated possibility of a subsequent estimate is used to acquire the noise signal from the possibility dissemination purpose. Since the determination of the concentrated probability estimate established on GGD cannot be reached directly, the Newton-Raphson law is used to acquire the steadfastness of the wavelet constants of a real signal. With respect to the signal noise proportion and the mean squared error, the outcomes show that the suggested technique can effectively eliminate the noise of the electromagnetic image, and that the implementation is much superior to the smooth sill technique. wavelet. The solid inception of the wavelet. [8]

(F. Huang *et al*., 2017) This is one of the best algorithms for image noise elimination through the non-local media resource (NLM) algorithm, superior ability to maintain image detail, widely used for image processing. image, remote sensing. However, the time involvedness of the algorithm is higher because of non-locality in the search for comparable pixels. As a consequence, the NLM algorithm cannot meet the requirements of some instantaneous applications. In this paper, we implemented an NLM of parallel algorithms based on the Intel Xeon processor's Phi Phi processors, which were developed and solved for this problem and equipped with the integrated Intel architecture (MIC). Although the parallel algorithm provided sufficient acceleration, the resulting acceleration showed a gradual dissemination for dissimilar image sizes. This consequence was not predicted centered on the speculative consideration that the acceleration should be self-regulating of the size of the input dataset. To solve this problem, I optimized the parallel algorithms by doing additional preprocessing and adding approaches to reduce the number of nested MIC cycles. Finally, the experiments were performed using customary and improved descriptions by using RS images of dissimilar sizes. Numerous assumptions can be acquired from the investigational outcomes: 1) the prevailing equivalent algorithm can achieve better acceleration with the PCM sound card, 2) optimized parallel algorithm; you can eradicate the progressive dissemination of full acceleration and processing of images of Significant RS. [9]

(A. Ertürk, 2017) Unmixing provides a summary of hyperspectral data and is useful for many imaging applications. Recently, hyperliterature has introduced the elimination of spectral separation and image noise. So far, however, only spectral information has been used to suppress noise based on the mixture. Most of the material termination methods found in the literature depend only on the spectral information, but Spatial Spectrum Pretreatment (SSPP), the last term, is placed in the most probable and homogeneous region. finite element extraction can be improved with the hypothesis. In this letter, it is proposed to use spectral resolution TPMSs to control the last element of the spatially uniform extraction field. By improving end-of-end suppression performance, noise elimination performance has been improved. In addition, SPP (the proposed approach continues and rare memory / finite anomalies that may contain important terminal elements such as stress cultures of rare minerals or military structures, spatial pretreatment may be lost by the inclusion of). For abstract or large compressed data, such abnormal rejection can have undesirable consequences. Therefore, the proposed approach offers better mixed-based noise suppression characteristics while maintaining extreme extremes. [10]

(E. Luo, S. H. Chan and T. Q. Nguyen, 2017) In this article, the author suggested that one of the best algorithms is an adaptive learning method for learning image-based image correction for removing images. The newest algorithm that is called Expectation Maximization (EM) variation, requires a nonspecific assumption erudite from a standard peripheral database and familiarizes it to the noisy picture to produce an explicit priority. Unlike obtainable techniques that syndicate ad hoc interior and exterior statistics, the suggested algorithm is obtained strictly using a hyper-a priori Bayesian viewpoint. There are two involvements from this document. Firstly, we suggest a complete beginning of the EM matching algorithm and determine techniques for improving computational convolution. Secondly, in the privation of the suppressed image, we demonstration in what way EM alteration can be changed based on prefiltration. The investigational outcome indicate that the suggested variation algorithm provides improved noise eradication outcomes than the one deprived of variation and greater to numerous advanced algorithms. [11]

(N. Riyahi-Alam *et al*., 2010) In this study, the highest plate-based probability blur (MPLE) rating was performed to eliminate the noise of SPECT images and was compared to other noise suppression techniques for example spraying or filtering. Butterworth. The Platelet-based MPLE fragmentation as a multi-scale disintegration method has already been projected for improved rendering of limits and surfaces due to Poisson noise and the intrinsic softness of such images. We apply this technique in computer-generated and realistic SPECT images. For NEMA ghost images, the level of noise measured earlier ($M_b$) and afterward ($M_a$) noise elimination using the platelet-based MPLE method were Mb = Ma = 0.1399. In the study of patients with 32 cardiac SPECT images, and the variance amongst the noise level and the SNR was earlier and afterwards the approach ($M_b$ = SNRb9.7762, $M_a$ = 0.7374, SNR = 4 1.0848). Therefore, we found the variance of the SNR coefficient (CV) for the images deleted by this algorithm compared to the Butterworth filter (145/33%). An SNR change was obtained for 32 SPECT images of the brain (196/17%). Our outcome show that based on Mple-Wafer, a valuable technique for removing SPECT images based on the most homogeneous image, better SNR, better targeting radiation absorption, and reducing the background activity of interfering radiation is the usual noise compared to other methods to eliminate. [12]

(M. Rosa-Zurera, A. M. Cóbreces-Álvarez, J. C. Nieto-Borge, 2007) In this paper, which is to decrease the speckle noise is started in that one of the main problems in the processing of imitation aperture radar (SAR) images to solve. This document describes a method for deleting a point and improving SAR images in a wavelet domain. In particular, we use edge detection in SAR images with the soft threshold method. One of the main objectives of the noise reduction process is to consider whether it is possible to dampen noise and at the same time keep sharp edges and shapes. Investigational outcomes on heterogeneous SAR images suggest that the proposed algorithm is technically suitable for this purpose and allows us to improve classification and recognition detection performance on the SAR basis. [13]

In this work, the author has an image search technique centered on the merger of graphically alike picture blocks is in the context of one or more images of the same suggested scene. The proposed approach takes into account the difference between the frame and the presence of atypical values by which they represent stirring substances in the passage extract. The main application is proposed by the suggested method for stabilizing the images of different images which combines the integration of the image with the effect of camera shake by a plurality of shorter exposed image scene frames. Because of its small consociate to distinctive surrounds are noisy, because they are by a blur movement less damaged than by a larger, open framework. The suggested technique is established by sequences of experimentations and evaluations. The outcomes demonstrate the credibility of the suggested technique to enhance image excellence by tumbling noise and mimicking coverage time. [14]

(F. Flitti, C. Collet and E. Slezak, 2007) The above approach was tested with true high-resolution multi-zone galaxy astrophysical images of the Hubble depth field on the Hubble- Space-Telescope at the 6 wave-lengths of the respite of the FUV at Band I (Fig.3). Using a pyramid algorithm, we perform 4-step wavelet transformations for each band. The fused image, which was finally reconstructed using different merging rules, is shown in FIG. Rules one and two obviously surpass rule three and the average of single bands. The constructed image summarizes the main features of the object in an image retrieved from the screen and identifies the overall structure of the galaxy. [15]

(C. Theys and H. Lantéri, 2006) In this article, the author claims that this is a new technology that makes it possible to buy data from Astrophysics L3CCD cameras to avoid reading noise due to the inclusion of conventional CCD. The physical process leading to the data was previously described by the density of "gamma-fish". We propose to discuss the model and obtain an iterative DE convolution algorithm for the data. Some simulation results are contained in synthetic astrophysical data, which shows the interest of L3CCD cameras in producing very low intensity images. [16]

c. State-of-the-art Image Denoising Techniques

In this research work, we have particular carried out denoising using following three state-of-the-art algorithms due to their tremendous performance.

1. NL-means

(A. Buades, B. Coll and J-M Morel, 2005) It is an algorithm that removes the noise of images called non-local media (NL averages) based on the average dose of all the pixels in the image. It is also the best algorithm for removing extra sound from images. Non-local media algorithms do not provide this assumption, but assume that the image contains a large amount of redundancy. [17]

2. K-SVD

(M. Elad and M. Aharon, 2006) This is the second state of the algorithm which must remove Gaussian noise and zero noise from zero to zero from the specified image. In the K-SVD algorithm, there is a dictionary that clearly shows the contents of the image. This happens with the most corrupted image, which is the image removal technique used exclusively in this project. K-SVD is a signal representation technique that can generate a dictionary that can execute arbitrary signals with scattered atom mixing from a series of signals. This algorithm always produces superior results in terms of PSNR and SSIM. [18]

3. BM3D

(K. Dabov, et al., 2007) This is one of the best noise suppression algorithms that can affect all types of images. BM3D is a new way to eliminate noise based on the fact that images are scattered local expressions in the field of transformation. This distribution is solved by collecting the same 2D images in a three-dimensional group. This document proposes an implementation of an open source method. We discuss all parameterization options and confirm actual optimization. The description method is rewritten in a new notation. Significant improvements have been obtained with conventional filters, especially Wiener. We will detail this new noise suppression strategy and its algorithm based on its effective implementation. There is also an add-in that removes noise from color images. Experimental results show that this algorithm can be computed using calculations that provide modern noise suppression performance in terms of peak-to-noise ratio and subjective visual quality. [19]

### III. DESIGN AND METHODOLOGY

This section is about the main exploration approach used in this project and provides an insight on the tools used for this project. The first section describes the focus of this research. It also provides a brief description of the project. The next section highlights the resources used in this project. Third section defines the programming language (MATLAB®) used for this project. It clearly specifies the language and all the tools within that package. Fourth section introduces to the data formats used within this project. The fifth section identifies the environment that was chosen for testing and experiments. It also highlights the pros and cons of each environment. Sixth section provides the main prototype or working model of the project. Seventh section provides the design considerations and reasons for choosing the specific techniques.

a. Technical Overview

Image denoising algorithm changed noised image with better form of image. Image get noised due to several reasons; such as while taking from defected camera, though sending to another storage device, reflects of lights while captioned from video. Noise is the corrupt signal that affects the original signal and thrown away pixel from original position. This is the caused that's why image gets blurred or degraded. Noise can disturb quality of digital or binary images. There are numerous potential sources of noise. The main reason of noisy signals are low quality camera instruments, defect in operation while sending to another storage medium, and natural lights effects or poor quality of machine;

To proceed, various random elements of Gaussian noise are added to an original clean image. The process will be divided into two major steps, followed by random noise being removed by three different image denoising algorithms. The objective of the image denoising processing is to enhance the possible facts for humanoid explanation. It is also dispensation image for storing broadcast and presentation for self-governing machine awareness. Digital images are often corrupted by impulse noise in transmission error. Different kinds of noise-corrupt an image during the process of acquisition, transmission, and reception, and storage and retrieval. All these factors combined and made an image noisy. Although, after that image needs to be restored before it is used. To complete the purpose denoising techniques are used. A lot of research work have been carried out on the restoration of images corrupted by AWGN. Image deploring and image de noising are the two sub areas of image restoration.

b. Recourses Required

1. Study Related Resources

- Understanding of and Programming Skills in MATLAB®
- Basic Language Programming Skills
- Matrix and Relevant Mathematics
- Graphs
- Review Research Papers of Image Processing

2. Other Resources Required

- Internet
- Greyscale Images of Different Types (PNG, JPG, TIFF, GIF)
- Computer Machine
- 64 bit Windows

c. Implementation Methods

This project will be implemented in by developing a MATLAB® Code; That will take Greyscale image as an input and calculate PSNR and SSIM of the image; than store the same results in graphs. In later step, images printed along with graphs that can show and compare image denoising algorithms.

d. Techniques

The following is a list of techniques that we have used to complete this project:
- Grayscale images collection
- MATLAB® code development
- Study and implementation of algorithms
- Corrupting clean images via AWGN
- Denoised images using different denoising algorithms
- Analyzing and comparing extensive set of results

e. Limitations of Technique

The main limitations of the techniques, which require future work, are detailed as below
- Simulation takes lot of time
- Large size mages put computational burden over the machine

f. Provided Resources

The results in our work are compiled by using MATLAB®, mathematical and graphical tools, basic programming techniques, and using already developed algorithms. This is also accompanied with collecting standard datasets used by image processing community. In the first task, we have collected 9 images datasets for testing. In the second task, we have collected 10 natural images used by researchers for same. In the third simulation process, we have collected 10 texture images, and at the last we have tested about 21 manmade images. All these images have been collected from internet resources where all these images were mostly used by image processing community.

g. Programming Language

There is no specific programming language used in this project, Since This is Research based project and required lot of simulations of MATLAB® which include basic programming skills, use of Loops and Structural Programming techniques. MATLAB® built-in functions are used widely in this project. A privately-owned programming-language advanced by Math Works, MATLAB® suggest matrix persuasions, conspiracy of purposes and facts, algorithm accomplishment of algorithms.

h. Image Denoising Techniques

In this project, we have compared three difference algorithms against an extensive set of images having following three different image sizes 64x64, 128x128, and 256x256, and over a vast range of noise levels.

1. BM3D

It is one the best denoising algorithm that can affect on all types of image. The new BM3D eliminates noise, as is the case with a saint representative. Distribution within a group of images is a sign of the group's 2D body. Only the source developer can be opened in this document. All of these options are available, the settings must match and must be true. This is the best I can confirm. The explanation has been rewritten in a new spelling. I hope this will become transparent to the original notation. At the end of the index, however, the difference between the specification and the original notation is displayed.

2. K-SVD

This is second state of art image denoising technique used this project. K-SVD is a method of representing a signal from a series of signals that can produce a dictionary that can approximate any signal with a combination of scattered atoms. This algorithm always gives optimum results in term of PSNR and SSIM.

3. NL-means

In addition, the best algorithm is considered to remove additional noise from the image. Non-local media algorithms do not provide this assumption, but they assume that the images contain a large amount of redundancy. Non-local media algorithms do not offer this assumption, but they assume that the images contain a large amount of redundancy. This redundancy can be used to eliminate sound images. This redundancy can be used to eliminate sound images. The non-local media source (NLM) algorithm is one of the best algorithmic eliminations of image noise to explain the representation of the cause image, and is widely used for remote sensing (RS). However, the time complexity of the algorithm is very high, because no similar pixel can be seen here. As a result, the NLM algorithm cannot meet the requirements of some real-time applications. To solve this problem, this work develops and implements the Intel Integrated Architecture (MIC) and hardware-based parallel NLM algorithm based on the Intel Xeon file.

i. Selection of Images

We have selected images from internet resources most commonly used by Image processing have been discussed and the results from various simulations and experimentations will be discussed in the upcoming sections.

community. The Images datasets are divided in four parameters. In first dataset, we have selected 9 standard greyscale images. In second dataset, we have selected 10 natural images. In third dataset, we have selected 10 texture images, and in the last, we have selected almost 20 manmade images.

j. Assessment Scenario

The assessment procedure in our work is divided into two different ways as given below:

1. Objective Assessment

This is mathematical measurement that judge the quality of image by value, i.e., expressed in terms of peaks signal-to-noise ratio (PSNR) and structure similarity index (SSIM).

2. Subjective Assessment

Sometime given PSNR/SSIM for denoised is not good or extra valued, Therefore, viewing by human eye, image result may be opposite to the exact PSNR/SSIM. This type of assessment is also acknowledged and we have provided the resultant images in the form of figures as well.

k. Process Flowchart

In Fig. 4, we have shown a flowchart of the entire workflow of our work. Firstly, we gather datasets which is then followed by extracting images from them. Afterwards, AWGN is added to the standard test images and the state-of-the-art algorithms are applied for denoising. Finally, the results are compared.

In the above section, different methods for implementation has been discussed. It also specifies the advantages and disadvantages of each technique. The methods tested for this thesis

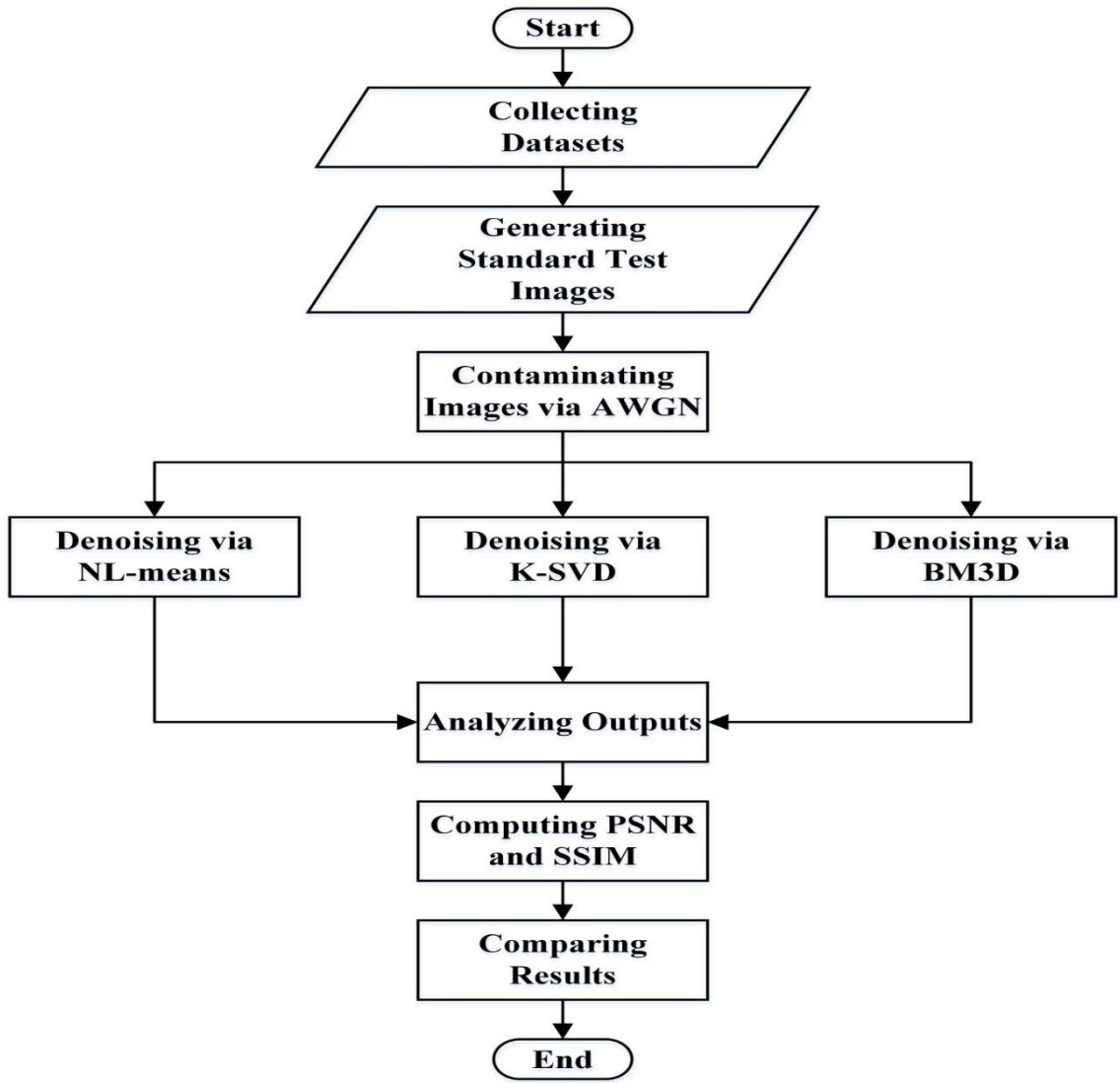

Figure 4: Flowchart of the proposed work

## IV. IMPLEMENTATION AND EXPERIMENTAL RESULTS

As far as implementation into the functional perspective is concerned, it is being implemented as denoising algorithm. In this section, we will discuss selected datasets, and the images therein, for this project. We will highlight the implementation results and the output figures will be shown over here.

a. Tools

To implement the project, following tools has been used in this project:

1. MATLAB®

2. Grayscale Images

3. Matrices and Graphs

b. Collecting Images

We have collected these images from publicly available online internet resources. These images are used by image denoising community very commonly. In this process, we have selected four types of datasets namely general grayscale images, natural images, texture images, and artificial images. A dataset, in our case, generally comprises of five to ten images of same types.

c. Generating Standard Test Images

After collecting images, we generate standard test images, in which first dataset consists of nine standard grayscale images, while the second dataset consists of six natural images. The third dataset consists of ten texture images. Lastly, we use 20 manmade artificial images.

d. Contaminating Images via AWGN

We have completed over 40 images for testing. In the first process, all images were corrupted by AWGN. Each image was converted into noisy before it is used or denoised.

e. Denoising via NL-means, K-SVD, and BM3D

Once we have a noisy image, this noisy image is supplied to the denoising algorithm. In particular, we used three different denoised algorithm namely NL-means, K-SVD, and BM3D, and their results were stored for further analysis.

f. Analyzing Output

The Images were analyzed by two different assessments namely subjective and objective assessments respectively. Objective assessment is taken on computation results whereas subjective assessment is recorded as human eye visualization.

g. Computing PSNR/SSIM

PSNR/SSIM was computing of each image and print on top of each image. More value of both will cause of better performance. The PSNR/SSIM was calculated by programming code.

h. Comparing Results

Since the project is based on analytical research, in this case we have made three tables in upcoming sections. This way, we provided an extremely efficient way to summarize the results in form of a more meaningful understanding.

i. Denoising Process

To test the denoising scenarios, we used a number of different images from different datasets. The self-explanatory results of implementation different denoising algorithms over a range of different images belonging to each dataset using simulated at different noise levels are given below. We present the detailed version of the MATLAB results from implementation in the form of graphs and figures from Fig. 5 to Fig. 14.

a. Results of Image Denoising

The tables 1-12 express the results of image denoising carried out using NL-means, K-SVD and BM3D. Three tables have been drawn for each dataset for a total of 4 datasets. This is because even for images belonging to one dataset, we apply the denoising over three different image sizes. We have noticed that if noise value of image is increasing, then PSNR/SSIM of the denoised image starts decreasing. In other words, noise value and PSNR/SSIM are inversely proportional. This is clearly shown in the provided tables.

In this research, we have tested over fifty images in three different sizes and results have been stored in in tables. Three image denoising techniques is compared where it is decisively declared that BM3D is currently one of the best algorithms for all types and sizes of images. NL-means is also effectively working on Images. K-SVD is just better than NL-means but cannot effectively outperform BM3D.

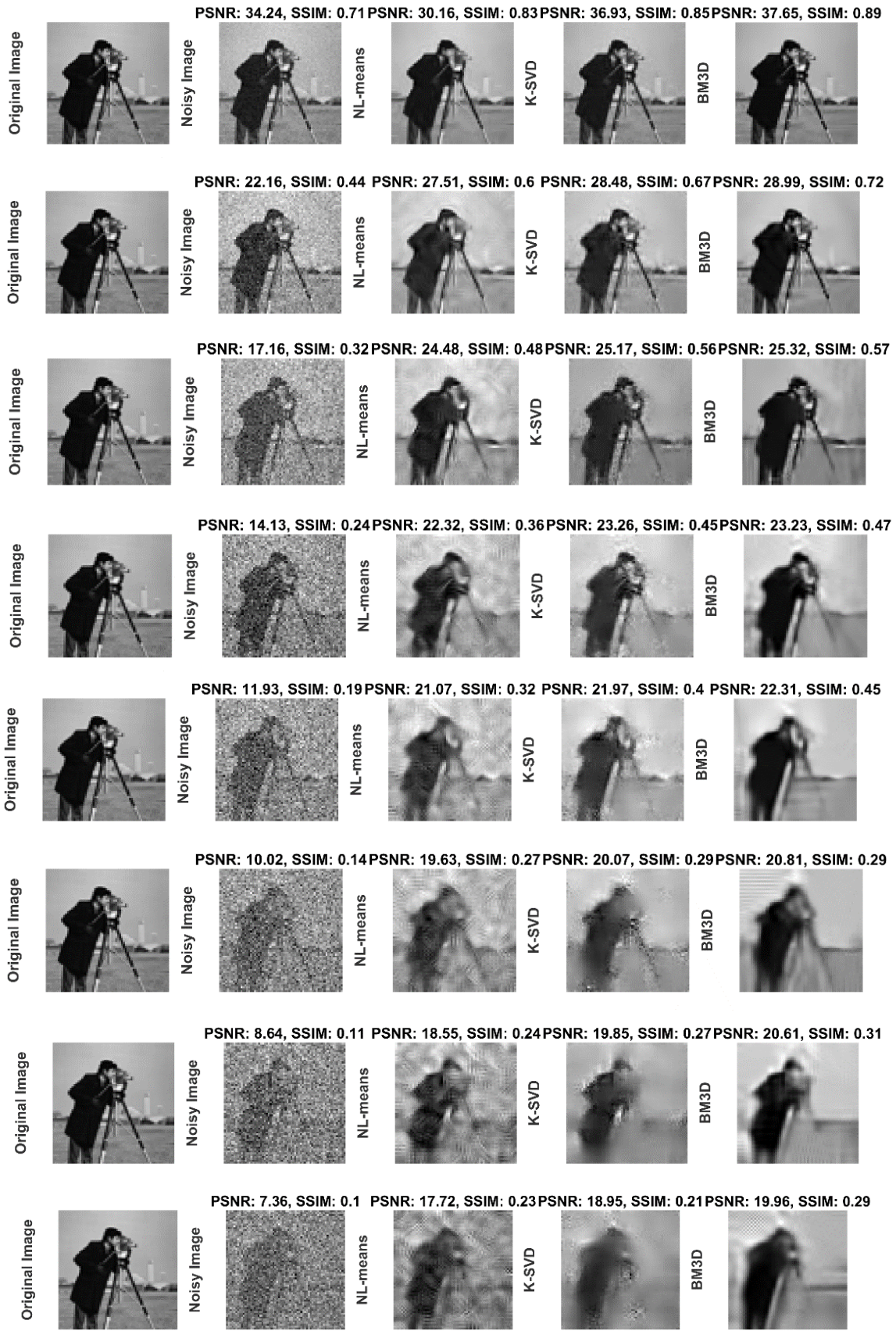

Figure 5: Denoising 64x64 test image from standard images dataset using 8 different noise levels. The PSNR, SSIM, and subjective results using NL-means, K-SVD, and BM3D algorithms are compared.

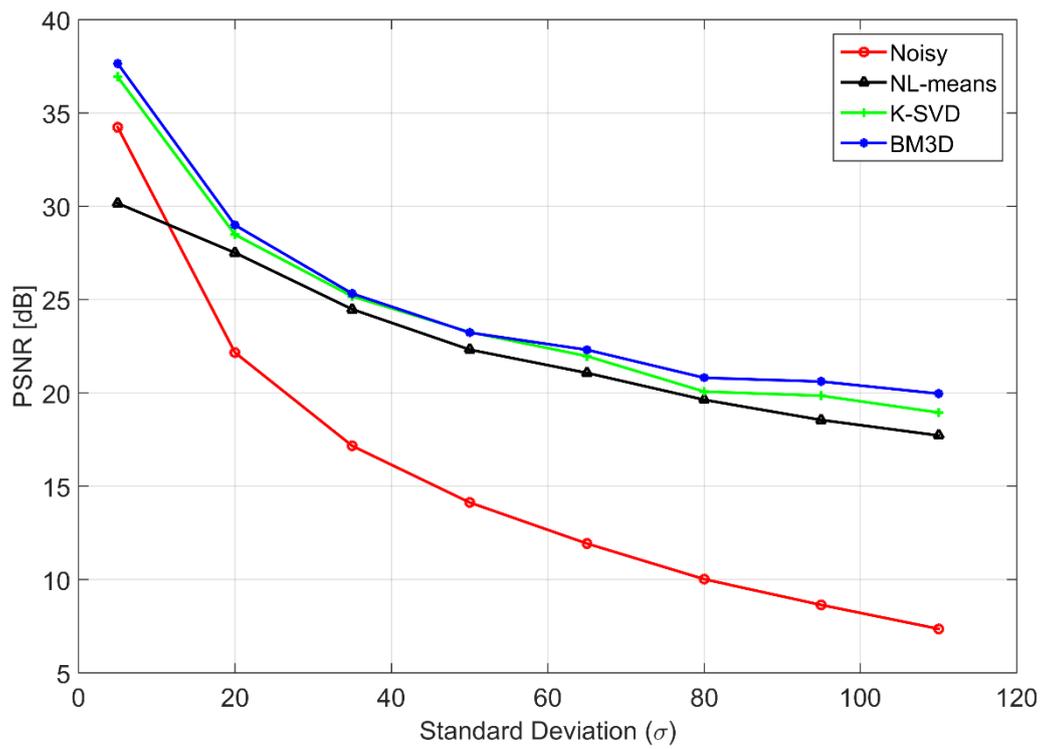
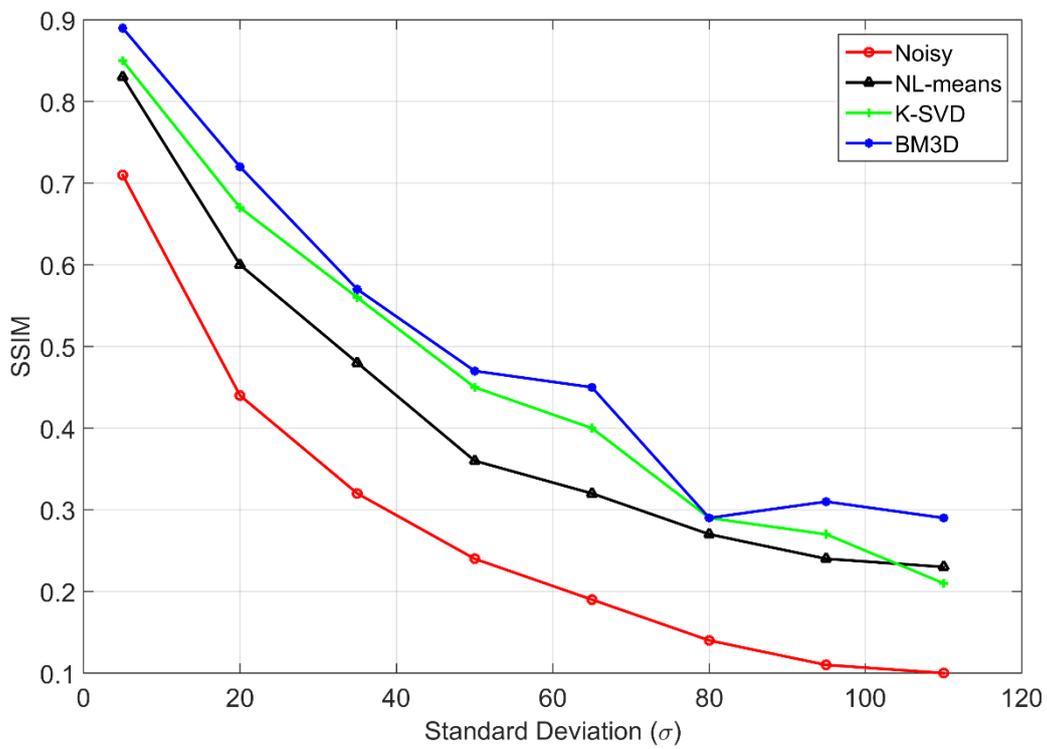

Figure 6: Graphical results of denoising 64x64 test image from standard images dataset using 8 different noise levels. The PSNR and SSIM, results using NL-means, K-SVD, and BM3D denoising algorithms are compared in the form of graphical illustrations.

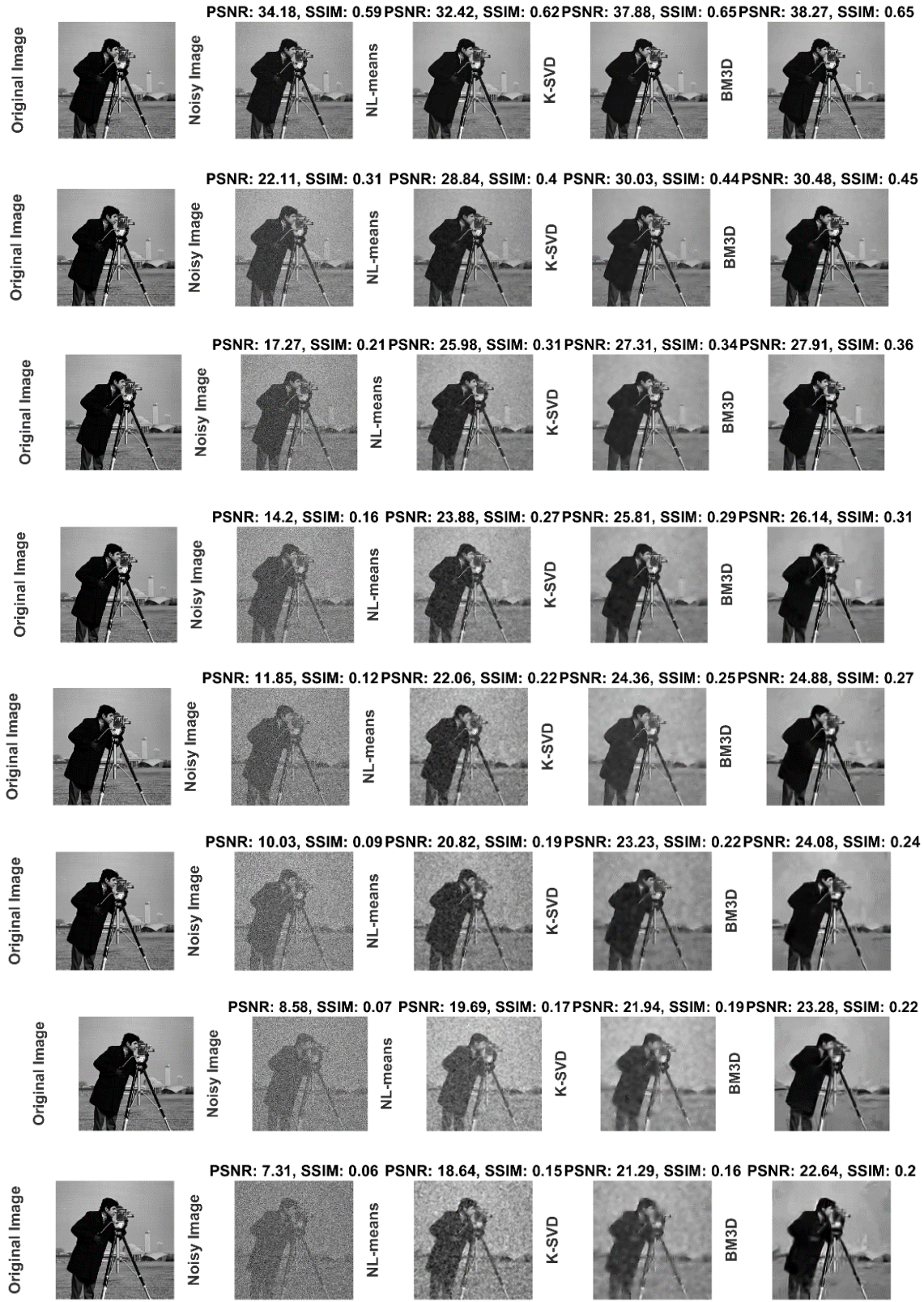

Figure 7: Denoising 256x256 test image from standard images dataset using 8 different noise levels. The PSNR, SSIM, and subjective results using NL-means, K-SVD, and BM3D algorithms are compared.

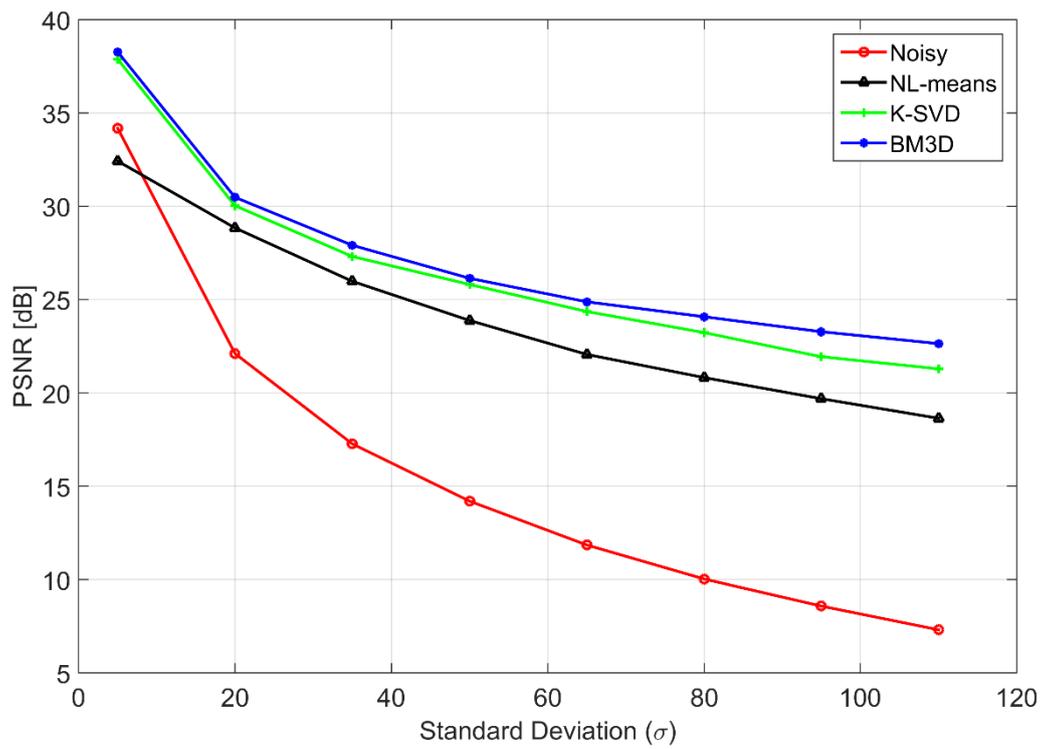
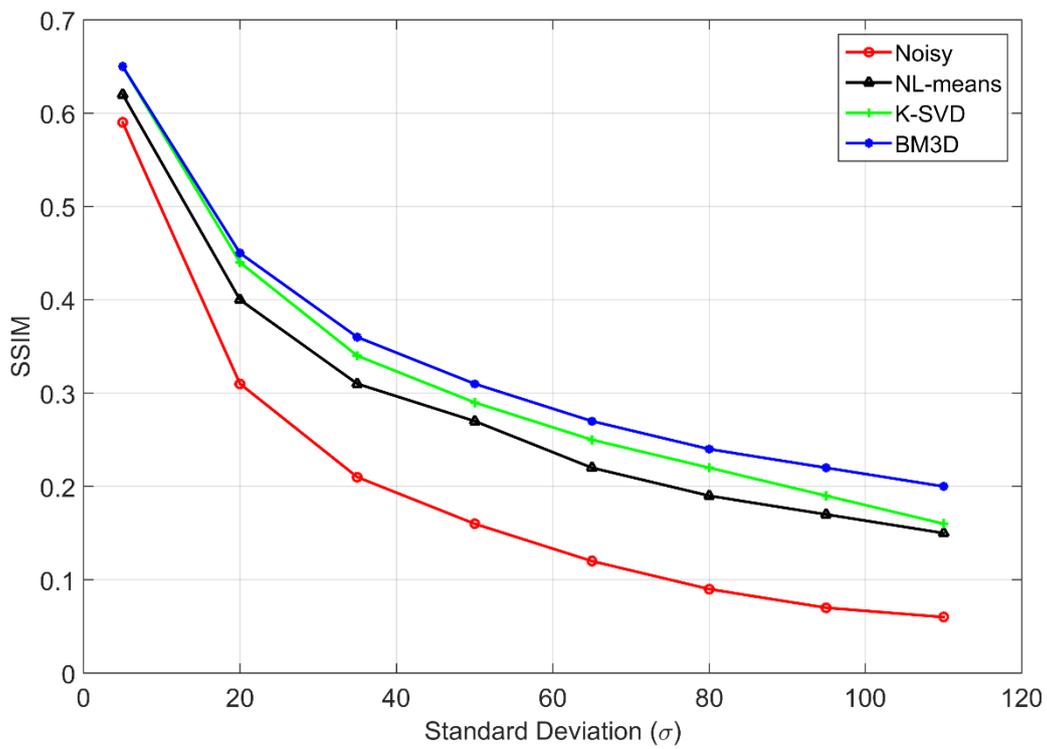

Figure 8: Graphical results of denoising 256x256 test image from standard images dataset using 8 different noise levels. The PSNR and SSIM, results using NL-means, K-SVD, and BM3D denoising algorithms are compared in the form of graphical illustrations.

\

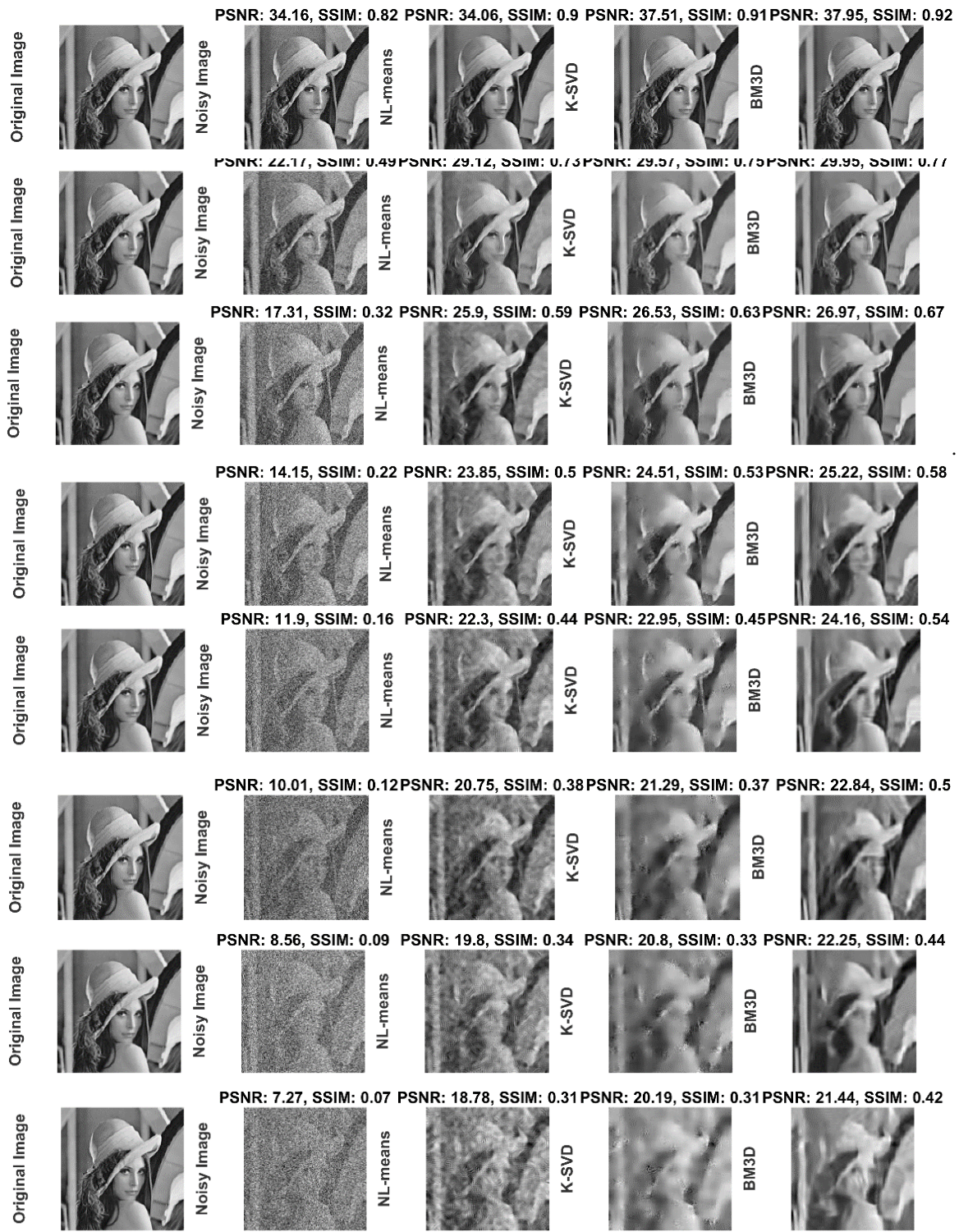

Figure 9: Denoising 128x128 test image from standard images dataset using 8 different noise levels. The PSNR, SSIM, and subjective results using NL-means, K-SVD, and BM3D algorithms are compared.

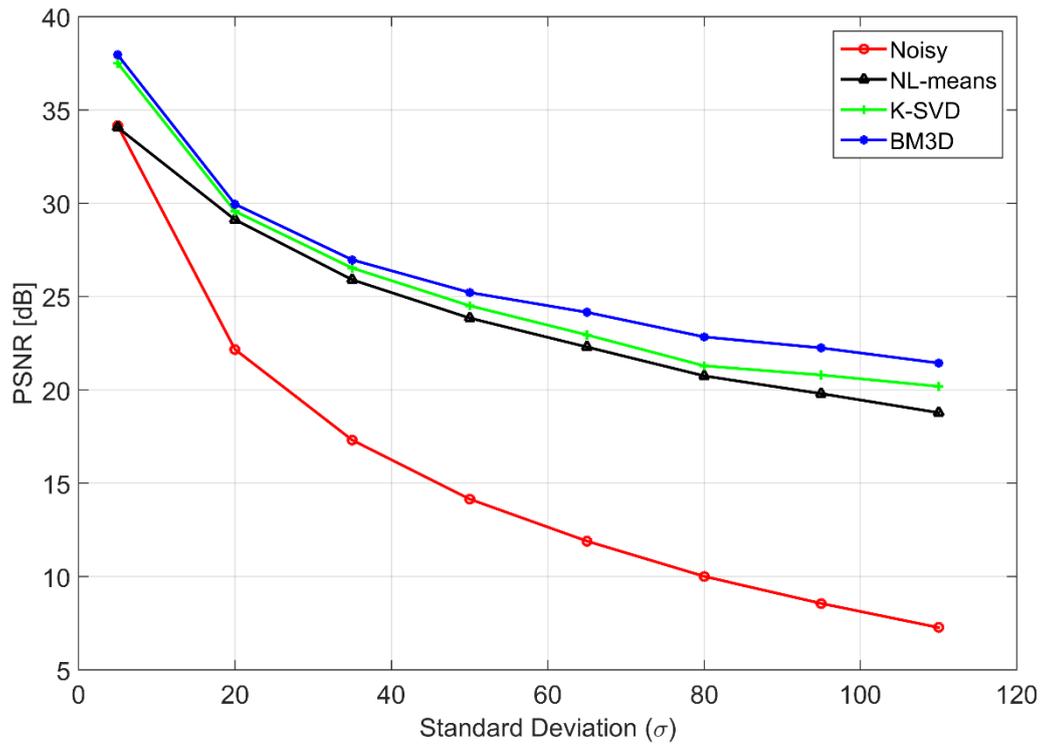

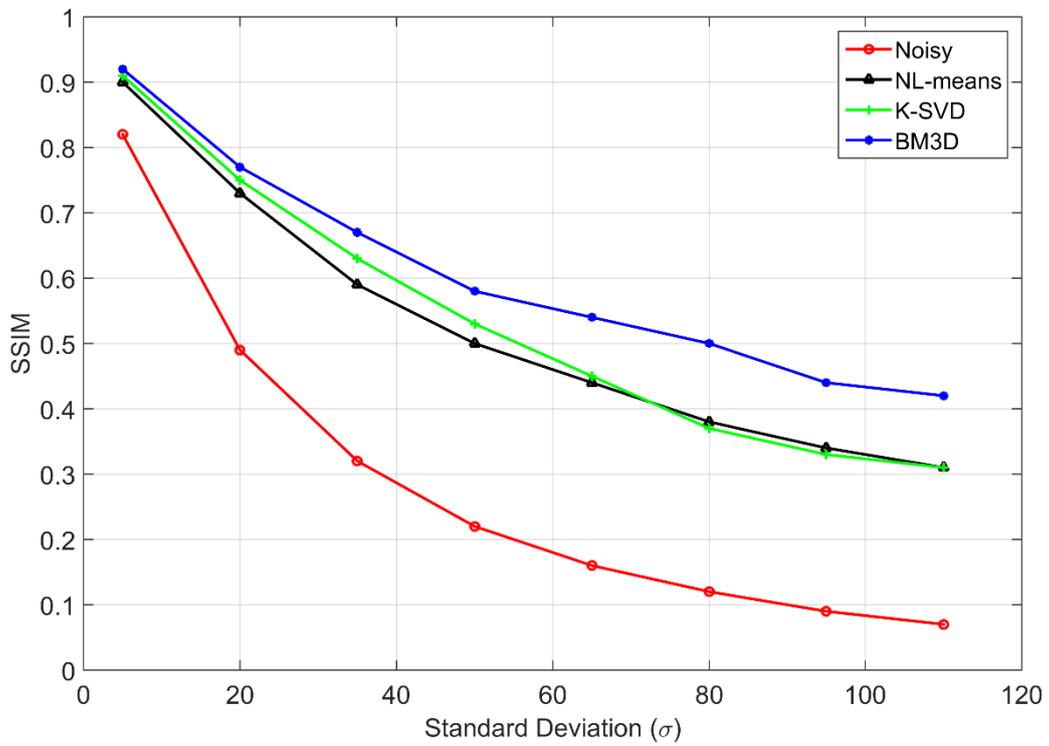

Figure 10: Graphical results of denoising 128x128 test image from standard images dataset using 8 different noise levels. The PSNR and SSIM, results using NL-means, K-SVD, and BM3D denoising algorithms are compared in the form of graphical illustrations.

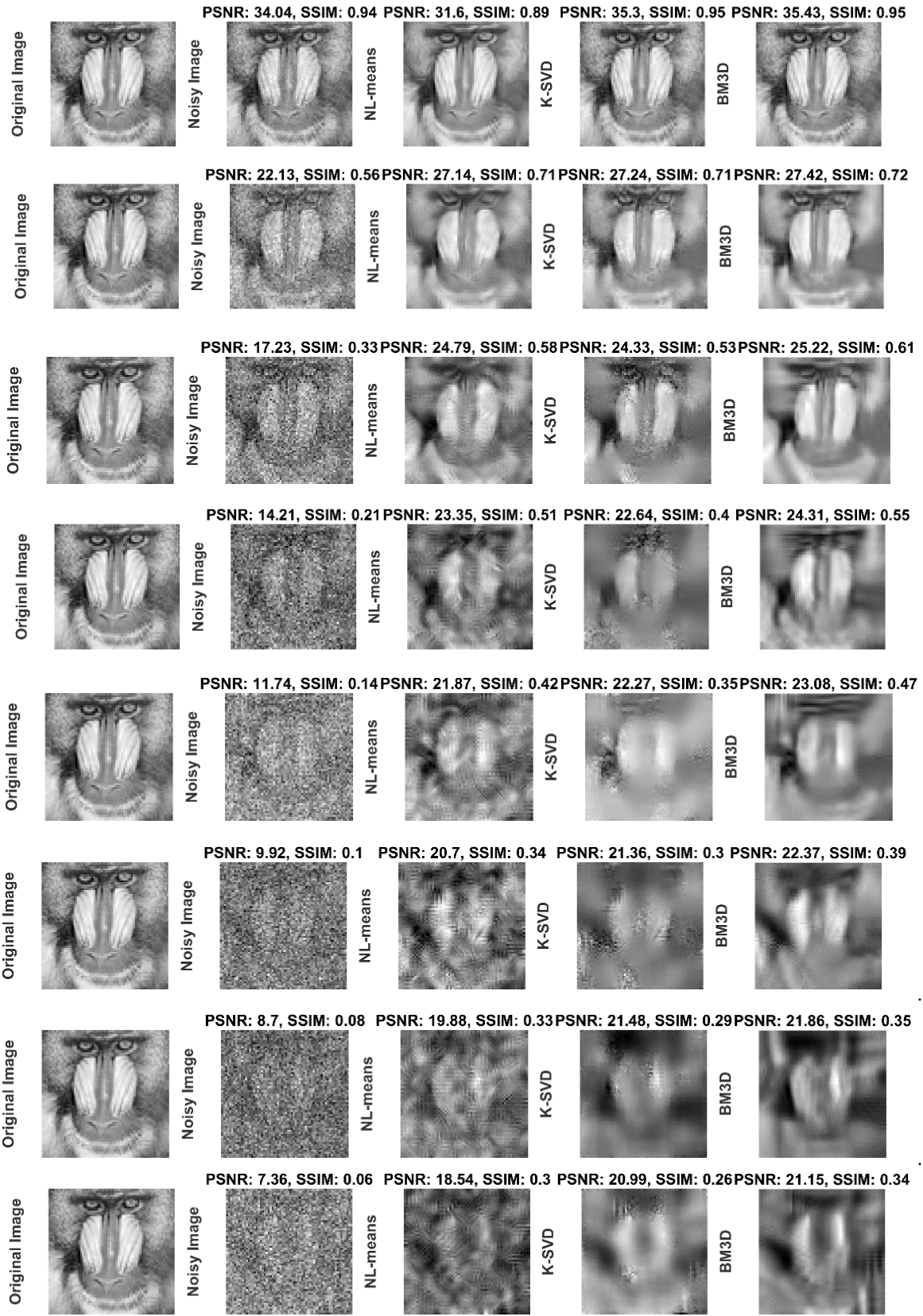

Figure 11: Denoising 64x64 test image from standard images dataset using 8 different noise levels. The PSNR, SSIM, and subjective results using NL-means, K-SVD, and BM3D algorithms are compared.

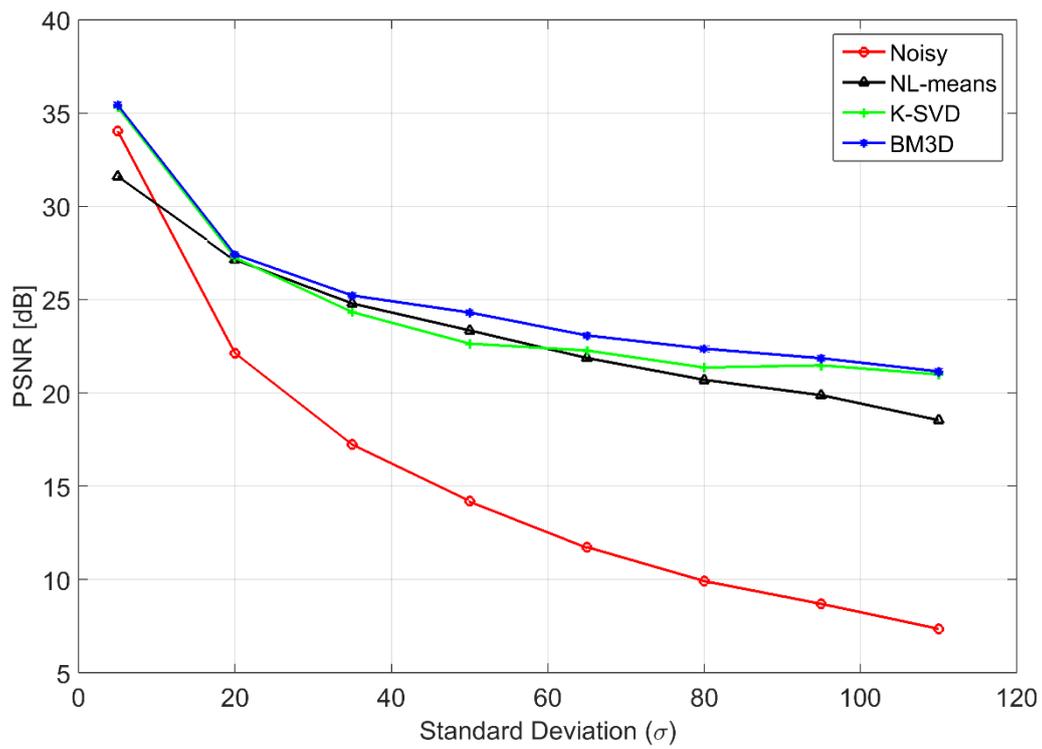
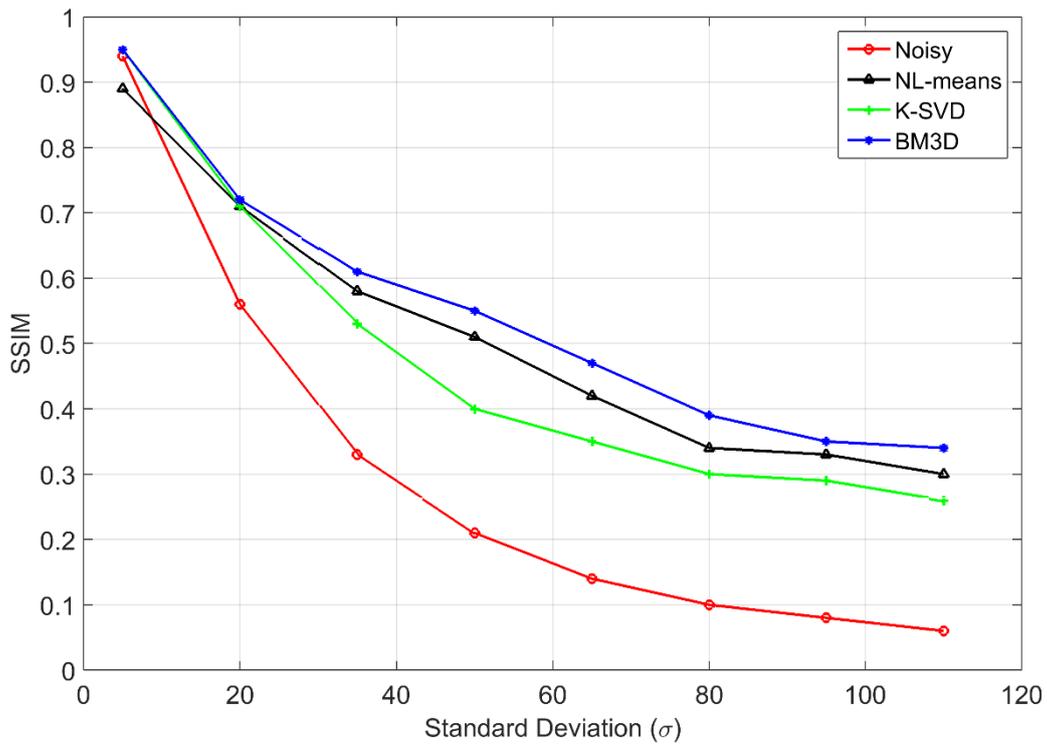

Figure 12: Graphical results of denoising 64x64 test image from standard images dataset using 8 different noise levels. The PSNR and SSIM, results using NL-means, K-SVD, and BM3D denoising algorithms are compared in the form of graphical illustrations.

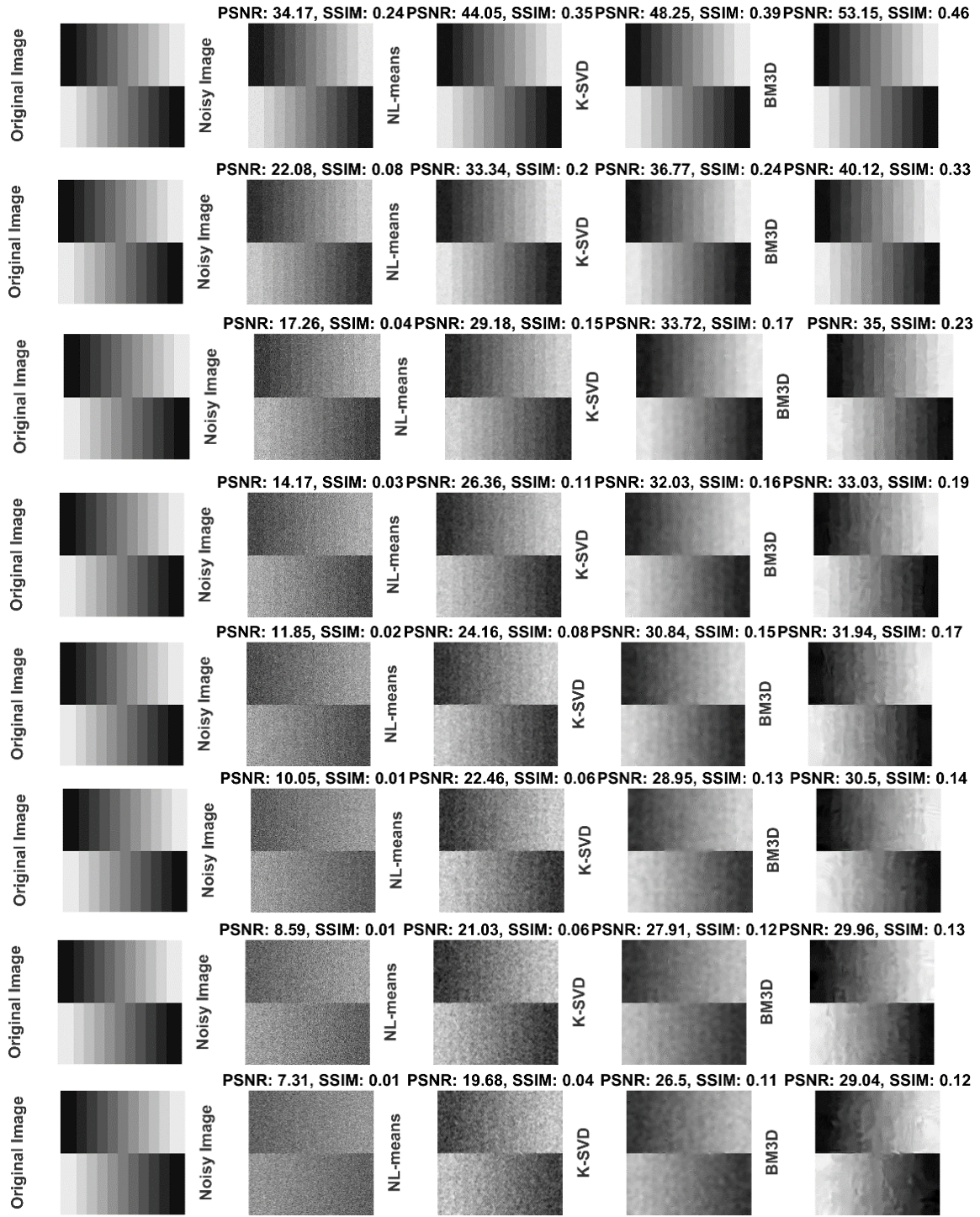

Figure 13: Denoising 256x256 test image from synthetic images dataset using 8 different noise levels. The PSNR, SSIM, and subjective results using NL-means, K-SVD, and BM3D algorithms are compared.

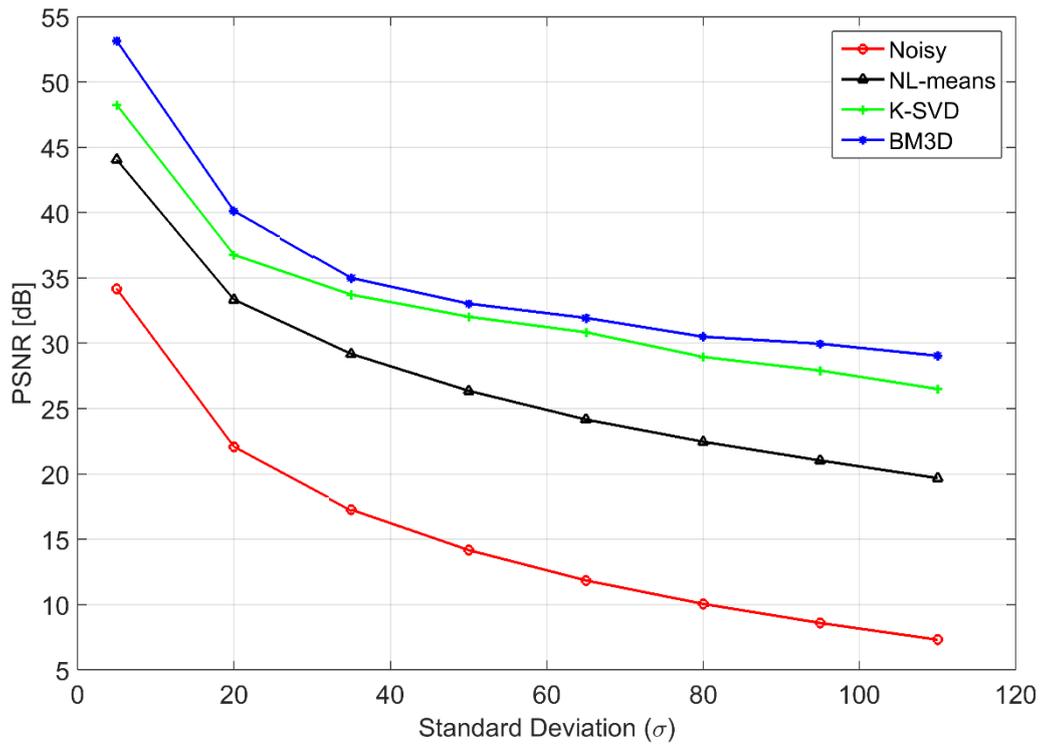

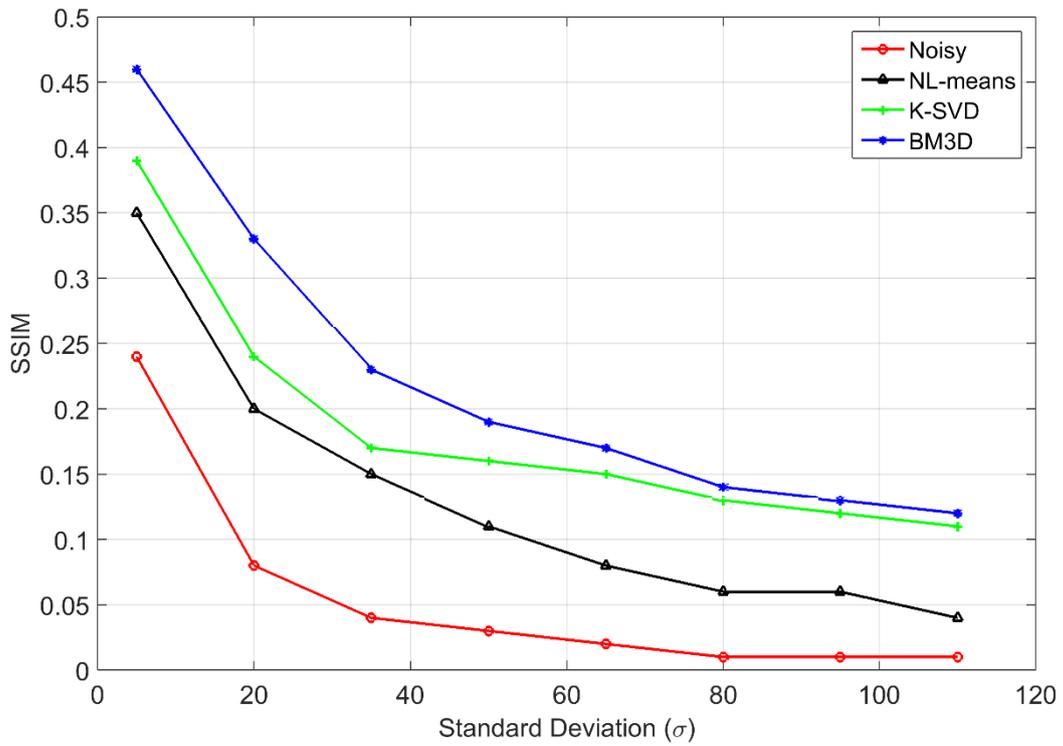

Figure 14: Graphical results of denoising 256x256 test image from synthetic images dataset using 8 different noise levels. The PSNR and SSIM, results using NL-means, K-SVD, and BM3D denoising algorithms are compared in the form of graphical illustrations.

Table 1: Denoising using NL-means, K-SVD and BM3D algorithms when applied on different *64x64* images from *Standard Test Dataset* using multiple noise levels

| Noise | Algorithm | Cameraman | Lena | Barbara | House | Peppers | Man | Livingroom | Boat | Mandrill |
|---|---|---|---|---|---|---|---|---|---|---|
| 5 | Noisy | 34.24/0.71 | 34.18/0.92 | 34.22/0.97 | 34.32/0.69 | 34.23/0.95 | 34.18/0.95 | 34.21/0.99 | 34.21/0.99 | 34.21/0.99 |
| | NL-means | 30.16/0.83 | 32.02/0.94 | 31.02/0.96 | 34.26/0.76 | 29.84/0.95 | 30.59/0.93 | 20.74/0.84 | 20.74/0.84 | 20.74/0.84 |
| | K-SVD | 36.93/0.85 | 36.24/0.95 | 35.61/0.98 | 37.97/0.77 | 35.84/0.97 | 35.19/0.96 | 34.24/0.99 | 34.24/0.99 | 34.24/0.99 |
| | BM3D | 37.65/0.89 | 36.54/0.96 | 35.82/0.98 | 38.86/0.79 | 36.05/0.97 | 35.39/0.96 | 34.25/0.99 | 34.25/0.99 | 34.25/0.99 |
| 20 | Noisy | 22.16/0.44 | 22.18/0.65 | 22.42/0.76 | 21.98/0.43 | 22.14/0.71 | 22.03/0.66 | 22.13/0.91 | 22.13/0.91 | 22.13/0.91 |
| | NL-means | 27.51/0.60 | 27.79/0.83 | 27.18/0.89 | 28.77/0.58 | 26.68/0.86 | 26.68/0.80 | 20.09/0.79 | 20.09/0.79 | 20.09/0.79 |
| | K-SVD | 28.48/0.67 | 27.66/0.84 | 26.86/0.89 | 29.44/0.60 | 26.98/0.87 | 26.64/0.81 | 23.32/0.92 | 23.32/0.92 | 23.32/0.92 |
| | BM3D | 28.99/0.72 | 27.94/0.85 | 27.16/0.90 | 29.46/0.60 | 27.03/0.87 | 26.84/0.82 | 23.18/0.92 | 23.18/0.92 | 23.18/0.92 |
| 35 | Noisy | 17.16/0.32 | 17.24/0.45 | 17.17/0.54 | 17.20/0.29 | 17.20/0.52 | 17.34/0.46 | 17.09/0.76 | 17.09/0.76 | 17.09/0.76 |
| | NL-means | 24.48/0.48 | 24.30/0.71 | 23.66/0.77 | 25.11/0.47 | 23.48/0.75 | 23.85/0.66 | 18.06/0.62 | 18.06/0.62 | 18.06/0.62 |
| | K-SVD | 25.17/0.56 | 24.41/0.72 | 23.43/0.77 | 26.02/0.52 | 23.75/0.77 | 23.60/0.65 | 19.89/0.80 | 19.89/0.80 | 19.89/0.80 |
| | BM3D | 25.32/0.57 | 24.90/0.76 | 23.64/0.78 | 26.49/0.55 | 23.74/0.77 | 24.03/0.69 | 19.23/0.76 | 19.23/0.76 | 19.23/0.76 |
| 50 | Noisy | 14.13/0.24 | 14.13/0.31 | 14.10/0.38 | 14.11/0.21 | 14.12/0.38 | 14.14/0.31 | 13.93/0.61 | 13.93/0.61 | 13.93/0.61 |
| | NL-means | 22.32/0.36 | 22.16/0.58 | 21.48/0.66 | 22.85/0.38 | 21.50/0.66 | 22.26/0.58 | 16.61/0.46 | 16.61/0.46 | 16.61/0.46 |
| | K-SVD | 23.26/0.45 | 21.82/0.55 | 20.93/0.61 | 23.56/0.40 | 21.40/0.64 | 21.84/0.51 | 17.75/0.65 | 17.75/0.65 | 17.75/0.65 |
| | BM3D | 23.23/0.47 | 22.90/0.65 | 21.92/0.70 | 24.20/0.45 | 21.83/0.68 | 22.43/0.57 | 17.60/0.60 | 17.60/0.60 | 17.60/0.60 |
| 65 | Noisy | 11.93/0.19 | 11.90/0.23 | 11.76/0.28 | 11.71/0.15 | 11.96/0.28 | 11.70/0.21 | 11.90/0.50 | 11.90/0.50 | 11.90/0.50 |
| | NL-means | 21.07/0.32 | 20.84/0.51 | 20.46/0.60 | 21.36/0.33 | 20.31/0.57 | 20.72/0.49 | 15.93/0.38 | 15.93/0.38 | 15.93/0.38 |
| | K-SVD | 21.97/0.40 | 20.59/0.46 | 19.97/0.54 | 22.02/0.33 | 20.18/0.53 | 20.35/0.41 | 16.67/0.51 | 16.67/0.51 | 16.67/0.51 |
| | BM3D | 22.31/0.45 | 21.31/0.55 | 21.02/0.63 | 23.13/0.40 | 20.72/0.59 | 21.14/0.49 | 16.34/0.42 | 16.34/0.42 | 16.34/0.42 |
| 80 | Noisy | 10.02/0.14 | 10.04/0.17 | 09.95/0.21 | 10.08/0.13 | 10.10/0.21 | 10.03/0.17 | 10.19/0.41 | 10.19/0.41 | 10.19/0.41 |
| | NL-means | 19.63/0.27 | 19.68/0.46 | 19.27/0.51 | 20.62/0.30 | 19.01/0.49 | 19.93/0.44 | 15.33/0.31 | 15.33/0.31 | 15.33/0.31 |
| | K-SVD | 20.07/0.29 | 19.20/0.36 | 18.81/0.43 | 21.30/0.25 | 18.34/0.39 | 19.89/0.36 | 15.63/0.35 | 15.63/0.35 | 15.63/0.35 |
| | BM3D | 20.81/0.29 | 20.63/0.52 | 19.96/0.55 | 22.71/0.38 | 19.67/0.54 | 20.61/0.44 | 15.50/0.27 | 15.50/0.27 | 15.50/0.27 |
| 95 | Noisy | 08.64/0.11 | 08.52/0.13 | 08.52/0.16 | 08.52/0.08 | 08.60/0.17 | 08.55/0.13 | 08.44/0.31 | 08.44/0.31 | 08.44/0.31 |
| | NL-means | 18.55/0.24 | 18.70/0.42 | 18.24/0.46 | 19.25/0.24 | 18.11/0.44 | 18.67/0.39 | 14.90/0.30 | 14.90/0.30 | 14.90/0.30 |
| | K-SVD | 19.85/0.27 | 18.86/0.34 | 18.07/0.34 | 20.16/0.19 | 17.77/0.32 | 18.96/0.29 | 15.21/0.29 | 15.21/0.29 | 15.21/0.29 |
| | BM3D | 20.61/0.31 | 20.16/0.49 | 19.18/0.49 | 20.70/0.25 | 18.79/0.45 | 19.38/0.37 | 15.33/0.24 | 15.33/0.24 | 15.33/0.24 |
| 100 | Noisy | 07.36/0.10 | 07.37/0.12 | 07.45/0.13 | 07.29/0.07 | 07.33/0.13 | 07.33/0.10 | 07.33/0.27 | 07.33/0.27 | 07.33/0.27 |
| | NL-means | 17.72/0.23 | 18.35/0.42 | 17.71/0.42 | 18.15/0.19 | 17.54/0.42 | 17.84/0.37 | 14.55/0.27 | 14.55/0.27 | 14.55/0.27 |
| | K-SVD | 18.95/0.21 | 18.53/0.33 | 17.43/0.25 | 20.09/0.15 | 17.41/0.31 | 18.74/0.26 | 14.96/0.21 | 14.96/0.21 | 14.96/0.21 |
| | BM3D | 19.96/0.29 | 19.74/0.50 | 18.89/0.45 | 20.82/0.24 | 18.31/0.41 | 19.28/0.38 | 14.96/0.17 | 14.96/0.17 | 14.96/0.17 |

Table 2: Denoising using NL-means, K-SVD and BM3D algorithms when applied on different *128x128* images from *Standard Test Dataset* using multiple noise levels

| Noise | Algorithm | Cameraman | Lena | Barbara | House | Peppers | Man | Livingroom | Boat | Mandrill |
|---|---|---|---|---|---|---|---|---|---|---|
| 5 | Noisy | 34.17/0.60 | 34.16/0.82 | 34.17/0.91 | 34.25/0.58 | 34.14/0.86 | 34.19/0.90 | 34.05/0.90 | 34.25/0.86 | 34.18/0.93 |
| | NL-means | 32.16/0.70 | 34.06/0.90 | 33.77/0.94 | 36.71/0.65 | 33.40/0.92 | 31.83/0.90 | 31.96/0.91 | 31.66/0.89 | 29.79/0.86 |
| | K-SVD | 37.80/0.74 | 37.51/0.91 | 36.93/0.96 | 39.33/0.68 | 37.38/0.93 | 35.91/0.93 | 36.33/0.94 | 36.33/0.92 | 35.11/0.95 |
| | BM3D | 38.39/0.77 | 37.95/0.92 | 37.25/0.96 | 39.89/0.67 | 37.75/0.94 | 36.15/0.94 | 36.61/0.95 | 36.57/0.93 | 35.24/0.95 |
| 20 | Noisy | 22.10/0.35 | 22.17/0.49 | 22.11/0.59 | 22.09/0.31 | 22.06/0.53 | 22.22/0.56 | 22.02/0.53 | 22.21/0.50 | 22.16/0.57 |
| | NL-means | 28.87/0.47 | 29.12/0.73 | 28.62/0.82 | 30.59/0.47 | 28.91/0.78 | 27.53/0.72 | 27.20/0.69 | 27.46/0.65 | 25.98/0.61 |
| | K-SVD | 29.82/0.54 | 29.57/0.75 | 28.72/0.82 | 31.92/0.50 | 29.12/0.80 | 27.63/0.72 | 27.78/0.72 | 28.10/0.66 | 26.58/0.66 |
| | BM3D | 30.26/0.55 | 29.95/0.77 | 29.12/0.84 | 32.74/0.51 | 29.43/0.82 | 27.85/0.73 | 28.03/0.76 | 28.16/0.69 | 26.53/0.65 |
| 35 | Noisy | 17.28/0.25 | 17.31/0.32 | 17.27/0.39 | 17.27/0.21 | 17.26/0.36 | 17.22/0.35 | 17.19/0.33 | 17.27/0.33 | 17.22/0.34 |
| | NL-means | 25.68/0.38 | 25.90/0.59 | 25.06/0.69 | 26.94/0.39 | 25.38/0.66 | 24.76/0.58 | 24.30/0.53 | 24.70/0.52 | 23.98/0.47 |
| | K-SVD | 26.87/0.44 | 26.53/0.63 | 25.42/0.70 | 28.96/0.44 | 25.83/0.69 | 24.96/0.57 | 24.73/0.54 | 25.27/0.53 | 24.26/0.46 |
| | BM3D | 27.25/0.46 | 26.97/0.67 | 25.82/0.74 | 29.91/0.46 | 26.23/0.72 | 25.34/0.61 | 25.16/0.61 | 25.47/0.56 | 24.40/0.48 |
| 50 | Noisy | 14.08/0.19 | 14.15/0.22 | 14.15/0.28 | 14.17/0.15 | 14.13/0.25 | 14.13/0.24 | 14.14/0.22 | 14.07/0.22 | 14.17/0.21 |
| | NL-means | 23.37/0.31 | 23.85/0.50 | 23.24/0.61 | 24.53/0.32 | 23.37/0.57 | 23.06/0.50 | 22.65/0.44 | 22.72/0.41 | 22.66/0.38 |
| | K-SVD | 24.77/0.37 | 24.51/0.53 | 23.26/0.60 | 26.44/0.36 | 23.78/0.60 | 23.23/0.46 | 23.09/0.42 | 23.45/0.42 | 22.85/0.32 |
| | BM3D | 25.08/0.39 | 25.22/0.58 | 24.12/0.66 | 27.60/0.40 | 24.22/0.64 | 23.79/0.52 | 23.56/0.49 | 23.73/0.44 | 23.36/0.37 |
| 65 | Noisy | 11.84/0.15 | 11.90/0.16 | 11.85/0.20 | 11.76/0.11 | 11.92/0.19 | 11.85/0.17 | 11.83/0.16 | 11.93/0.16 | 11.82/0.15 |
| | NL-means | 21.82/0.27 | 22.30/0.44 | 21.91/0.53 | 22.54/0.27 | 21.87/0.50 | 21.68/0.43 | 21.37/0.37 | 21.45/0.36 | 21.45/0.32 |
| | K-SVD | 23.50/0.31 | 22.95/0.45 | 21.79/0.51 | 23.98/0.27 | 22.04/0.51 | 21.85/0.37 | 22.03/0.34 | 22.33/0.36 | 22.06/0.27 |
| | BM3D | 24.02/0.34 | 24.16/0.54 | 23.08/0.62 | 25.81/0.35 | 23.10/0.57 | 22.61/0.45 | 22.79/0.43 | 22.82/0.40 | 22.72/0.33 |
| 80 | Noisy | 10.16/0.11 | 10.01/0.12 | 10.10/0.15 | 10.09/0.08 | 10.03/0.14 | 10.09/0.12 | 09.95/0.11 | 10.11/0.12 | 10.09/0.10 |
| | NL-means | 20.52/0.24 | 20.75/0.38 | 20.45/0.47 | 21.17/0.24 | 20.42/0.44 | 20.54/0.36 | 20.24/0.33 | 20.26/0.31 | 20.44/0.28 |
| | K-SVD | 21.95/0.26 | 21.29/0.37 | 20.52/0.42 | 22.95/0.23 | 20.58/0.44 | 20.89/0.30 | 21.01/0.30 | 21.17/0.27 | 21.62/0.25 |
| | BM3D | 23.14/0.32 | 22.84/0.50 | 21.81/0.54 | 24.54/0.31 | 21.95/0.53 | 22.01/0.39 | 21.80/0.37 | 22.06/0.34 | 22.18/0.30 |
| 95 | Noisy | 08.59/0.09 | 8.56/0.09 | 08.50/0.11 | 08.57/0.06 | 08.52/0.11 | 08.60/0.09 | 08.53/0.09 | 08.59/0.09 | 08.52/0.07 |
| | NL-means | 19.45/0.21 | 19.80/0.34 | 19.37/0.42 | 19.98/0.21 | 19.25/0.37 | 19.72/0.33 | 19.29/0.28 | 19.23/0.26 | 19.28/0.23 |
| | K-SVD | 21.05/0.20 | 20.80/0.33 | 19.96/0.39 | 22.15/0.21 | 20.00/0.38 | 20.63/0.29 | 20.68/0.25 | 20.35/0.24 | 21.13/0.21 |
| | BM3D | 22.32/0.28 | 22.25/0.44 | 21.00/0.49 | 24.04/0.29 | 21.14/0.46 | 21.65/0.37 | 21.37/0.32 | 21.43/0.30 | 21.46/0.24 |
| 100 | Noisy | 07.32/0.70 | 07.27/0.07 | 07.30/0.09 | 07.28/0.05 | 07.37/0.09 | 07.31/0.07 | 07.30/0.07 | 07.33/0.07 | 07.32/0.06 |
| | NL-means | 18.41/0.19 | 18.78/0.31 | 18.65/0.37 | 18.85/0.19 | 18.52/0.33 | 18.65/0.29 | 18.49/0.25 | 18.55/0.23 | 18.51/0.21 |
| | K-SVD | 20.22/0.18 | 20.19/0.31 | 19.38/0.34 | 20.99/0.15 | 19.16/0.34 | 20.17/0.26 | 20.05/0.22 | 20.16/0.24 | 21.23/0.21 |
| | BM3D | 21.60/0.25 | 21.44/0.42 | 20.61/0.46 | 22.66/0.24 | 20.50/0.44 | 20.91/0.32 | 20.84/0.27 | 21.32/0.28 | 21.26/0.22 |

Table 3: Denoising using NL-means, K-SVD and BM3D algorithms when applied on different *256x256 Standard Test Images Dataset* using multiple noise levels

| Noise | Algorithm | Cameraman | Lena | Barbara | House | Peppers | Man | Livingroom | Boat | Mandrill |
|---|---|---|---|---|---|---|---|---|---|---|
| 5 | Noisy | 34.18/0.59 | 34.12/0.75 | 34.14/0.86 | 34.09/0.58 | 34.18/0.75 | 34.25/0.95 | 34.18/0.89 | 34.13/0.83 | 34.19/0.95 |
| | NL-means | 32.42/0.62 | 33.94/0.78 | 32.09/0.88 | 37.48/0.61 | 35.02/0.81 | 30.77/0.93 | 30.81/0.87 | 30.86/0.83 | 26.44/0.84 |
| | K-SVD | 37.88/0.65 | 37.27/0.82 | 36.57/0.91 | 39.28/0.67 | 37.79/0.84 | 35.28/0.96 | 36.09/0.92 | 36.13/0.87 | 34.79/0.95 |
| | BM3D | 38.27/0.65 | 37.51/0.80 | 36.80/0.91 | 39.78/0.65 | 38.08/0.83 | 35.46/0.97 | 36.34/0.92 | 36.33/0.87 | 34.86/0.95 |
| 20 | Noisy | 22.11/0.31 | 22.16/0.37 | 22.10/0.54 | 22.09/0.25 | 22.13/0.37 | 22.19/0.67 | 22.09/0.52 | 22.13/0.48 | 22.14/0.65 |
| | NL-means | 28.84/0.40 | 29.45/0.59 | 27.87/0.72 | 31.43/0.38 | 30.18/0.64 | 26.68/0.79 | 26.84/0.64 | 27.26/0.60 | 23.80/0.62 |
| | K-SVD | 30.03/0.44 | 30.02/0.60 | 28.55/0.75 | 33.17/0.40 | 30.81/0.66 | 26.60/0.80 | 27.69/0.66 | 28.03/0.63 | 25.33/0.69 |
| | BM3D | 30.48/0.45 | 30.44/0.62 | 29.11/0.78 | 33.87/0.41 | 31.30/0.69 | 26.71/0.81 | 28.07/0.70 | 28.21/0.65 | 25.27/0.70 |
| 35 | Noisy | 17.27/0.21 | 17.27/0.23 | 17.26/0.36 | 17.25/0.15 | 17.24/0.24 | 17.21/0.45 | 17.23/0.33 | 17.29/0.31 | 17.25/0.44 |
| | NL-means | 25.98/0.31 | 26.44/0.46 | 24.73/0.58 | 27.72/0.30 | 26.87/0.52 | 23.70/0.66 | 24.02/0.49 | 24.49/0.46 | 21.73/0.45 |
| | K-SVD | 27.31/0.34 | 27.44/0.50 | 25.62/0.62 | 30.37/0.32 | 28.03/0.58 | 23.55/0.65 | 24.77/0.50 | 25.39/0.48 | 22.48/0.48 |
| | BM3D | 27.91/0.36 | 27.94/0.53 | 26.13/0.66 | 31.51/0.34 | 28.57/0.60 | 23.95/0.68 | 25.30/0.58 | 25.57/0.52 | 22.31/0.49 |
| 50 | Noisy | 14.20/0.16 | 14.13/0.15 | 14.12/0.25 | 14.16/0.10 | 14.15/0.16 | 14.15/0.31 | 14.13/0.22 | 14.20/0.21 | 14.18/0.31 |
| | NL-means | 23.88/0.27 | 24.26/0.38 | 22.70/0.48 | 25.25/0.24 | 24.46/0.43 | 22.18/0.56 | 22.31/0.38 | 22.74/0.37 | 20.61/0.35 |
| | K-SVD | 25.81/0.29 | 25.51/0.43 | 23.66/0.52 | 28.13/0.27 | 26.04/0.51 | 21.61/0.50 | 23.06/0.38 | 23.79/0.39 | 21.03/0.33 |
| | BM3D | 26.14/0.31 | 26.27/0.47 | 24.38/0.56 | 29.80/0.31 | 26.60/0.53 | 22.38/0.57 | 23.45/0.44 | 23.87/0.42 | 20.93/0.33 |
| 65 | Noisy | 11.85/0.12 | 11.84/0.11 | 11.88/0.18 | 11.87/0.07 | 11.85/0.12 | 12.01/0.23 | 11.87/0.16 | 11.88/0.16 | 11.86/0.22 |
| | NL-means | 22.06/0.22 | 22.68/0.32 | 21.32/0.41 | 23.52/0.21 | 22.73/0.37 | 20.90/0.49 | 21.11/0.32 | 21.26/0.31 | 19.64/0.30 |
| | K-SVD | 24.36/0.25 | 24.18/0.37 | 22.11/0.42 | 26.31/0.23 | 24.44/0.45 | 20.45/0.39 | 21.97/0.30 | 22.49/0.32 | 20.17/0.24 |
| | BM3D | 24.88/0.27 | 25.28/0.43 | 23.17/0.50 | 28.67/0.28 | 25.40/0.49 | 21.12/0.48 | 22.52/0.37 | 22.75/0.35 | 20.27/0.26 |
| 80 | Noisy | 10.03/0.09 | 10.06/0.08 | 10.05/0.14 | 10.10/0.05 | 10.09/0.09 | 10.03/0.16 | 10.10/0.12 | 10.04/0.11 | 10.09/0.16 |
| | NL-means | 20.82/0.19 | 21.40/0.28 | 20.18/0.36 | 21.93/0.18 | 21.31/0.32 | 19.71/0.41 | 20.07/0.27 | 20.13/0.26 | 18.93/0.25 |
| | K-SVD | 23.23/0.22 | 23.36/0.33 | 21.07/0.36 | 24.82/0.19 | 23.15/0.40 | 19.76/0.32 | 21.20/0.25 | 21.49/0.27 | 19.73/0.19 |
| | BM3D | 24.08/0.24 | 24.56/0.39 | 22.23/0.44 | 27.27/0.26 | 24.39/0.45 | 20.38/0.38 | 21.76/0.31 | 22.00/0.30 | 19.92/0.22 |
| 95 | Noisy | 08.58/0.07 | 08.57/0.06 | 08.58/0.11 | 08.60/0.04 | 08.61/0.07 | 08.39/0.12 | 08.59/0.09 | 08.58/0.09 | 08.58/0.13 |
| | NL-means | 19.69/0.17 | 20.12/0.24 | 19.19/0.31 | 20.47/0.16 | 20.04/0.28 | 18.86/0.38 | 19.16/0.24 | 19.24/0.23 | 18.25/0.23 |
| | K-SVD | 21.94/0.19 | 22.43/0.29 | 20.36/0.31 | 23.76/0.16 | 22.09/0.37 | 19.15/0.29 | 20.76/0.22 | 20.84/0.23 | 19.49/0.17 |
| | BM3D | 23.28/0.22 | 23.52/0.34 | 21.53/0.40 | 25.73/0.23 | 23.61/0.42 | 19.88/0.38 | 21.35/0.28 | 21.51/0.27 | 19.75/0.20 |
| 100 | Noisy | 07.31/0.06 | 07.29/0.05 | 07.31/0.08 | 07.32/0.03 | 07.30/0.06 | 07.19/0.10 | 07.35/0.07 | 07.29/0.07 | 07.31/0.10 |
| | NL-means | 18.64/0.15 | 19.04/0.21 | 18.27/0.27 | 19.55/0.14 | 18.91/0.24 | 17.95/0.35 | 18.35/0.21 | 18.26/0.20 | 17.50/0.20 |
| | K-SVD | 21.29/0.16 | 21.68/0.26 | 19.88/0.28 | 23.17/0.15 | 21.26/0.34 | 19.11/0.31 | 20.35/0.20 | 20.39/0.21 | 19.16/0.15 |
| | BM3D | 22.64/0.20 | 22.99/0.32 | 20.89/0.36 | 25.42/0.22 | 22.57/0.38 | 19.61/0.36 | 20.92/0.25 | 21.00/0.23 | 19.43/0.17 |

Table 4: Denoising using NL-means, K-SVD and BM3D algorithms when applied on different *64x64* images from *Natural Images Dataset* using multiple noise levels

| Noise | Algorithm | Bee | Bird | Boat | Bridge | Buildings | Cart | House | Owl | Terrain | Tomb | Water |
|---|---|---|---|---|---|---|---|---|---|---|---|---|
| 5 | Noisy | 34.33/0.88 | 34.25/0.83 | 34.13/0.92 | 34.14/0.96 | 34.18/0.88 | 34.04/0.95 | 34.10/0.90 | 34.25/0.95 | 34.19/0.95 | 34.19/0.95 | 34.16/0.87 |
| | NL-means | 33.39/0.89 | 32.72/0.89 | 30.58/0.91 | 29.57/0.91 | 35.34/0.92 | 29.61/0.93 | 31.13/0.84 | 29.10/0.89 | 27.90/0.84 | 27.90/0.84 | 32.00/0.86 |
| | K-SVD | 36.59/0.92 | 37.01/0.90 | 35.79/0.94 | 35.07/0.97 | 36.93/0.93 | 35.19/0.97 | 35.57/0.91 | 35.03/0.96 | 34.65/0.96 | 34.65/0.96 | 36.07/0.92 |
| | BM3D | 37.01/0.92 | 37.55/0.92 | 36.02/0.95 | 35.06/0.97 | 37.40/0.94 | 35.51/0.97 | 35.71/0.89 | 35.11/0.96 | 34.77/0.96 | 34.77/0.96 | 36.18/0.92 |
| 20 | Noisy | 22.25/0.54 | 22.05/0.46 | 22.07/0.55 | 22.20/0.68 | 22.13/0.44 | 22.19/0.69 | 21.97/0.56 | 22.08/0.65 | 22.15/0.62 | 22.15/0.62 | 22.09/0.48 |
| | NL-means | 28.80/0.75 | 28.17/0.67 | 27.04/0.69 | 25.73/0.75 | 28.87/0.65 | 25.82/0.78 | 27.10/0.70 | 25.59/0.71 | 24.64/0.59 | 24.64/0.59 | 27.01/0.58 |
| | K-SVD | 28.86/0.75 | 28.44/0.70 | 27.43/0.69 | 26.40/0.79 | 28.73/0.66 | 26.33/0.81 | 27.66/0.72 | 26.18/0.74 | 25.45/0.67 | 25.45/0.67 | 27.78/0.66 |
| | BM3D | 29.17/0.76 | 28.71/0.72 | 27.74/0.73 | 26.56/0.79 | 29.83/0.73 | 26.57/0.83 | 27.91/0.73 | 26.29/0.75 | 25.04/0.62 | 25.04/0.62 | 27.72/0.65 |
| 35 | Noisy | 17.29/0.34 | 17.35/0.29 | 17.29/0.36 | 17.34/0.46 | 17.29/0.24 | 17.32/0.48 | 17.06/0.35 | 17.30/0.43 | 17.30/0.40 | 17.30/0.40 | 17.42/0.29 |
| | NL-means | 25.20/0.61 | 25.81/0.55 | 23.97/0.54 | 23.49/0.63 | 26.19/0.52 | 22.86/0.61 | 24.34/0.54 | 23.14/0.55 | 22.81/0.46 | 22.81/0.46 | 25.07/0.46 |
| | K-SVD | 25.36/0.61 | 25.68/0.55 | 24.24/0.50 | 23.60/0.62 | 25.94/0.47 | 23.20/0.66 | 24.52/0.55 | 23.16/0.55 | 22.89/0.45 | 22.89/0.45 | 25.01/0.46 |
| | BM3D | 25.73/0.64 | 26.26/0.58 | 24.69/0.57 | 23.80/0.65 | 27.24/0.60 | 23.50/0.69 | 24.97/0.58 | 23.42/0.56 | 22.69/0.42 | 22.69/0.42 | 25.81/0.51 |
| 50 | Noisy | 14.03/0.23 | 14.20/0.20 | 14.26/0.25 | 14.06/0.31 | 14.05/0.13 | 14.30/0.34 | 14.15/0.25 | 14.20/0.30 | 14.13/0.25 | 14.13/0.25 | 14.09/0.16 |
| | NL-means | 23.23/0.53 | 23.42/0.43 | 22.38/0.47 | 21.54/0.51 | 23.84/0.38 | 21.18/0.49 | 22.60/0.45 | 21.72/0.46 | 21.67/0.37 | 21.67/0.37 | 23.53/0.38 |
| | K-SVD | 23.47/0.52 | 23.82/0.44 | 22.92/0.42 | 21.51/0.48 | 24.30/0.32 | 21.28/0.48 | 22.64/0.43 | 21.66/0.42 | 21.76/0.29 | 21.76/0.29 | 23.38/0.31 |
| | BM3D | 23.86/0.55 | 24.43/0.49 | 23.00/0.46 | 22.23/0.56 | 25.26/0.44 | 21.69/0.55 | 23.52/0.49 | 22.08/0.44 | 21.81/0.29 | 21.81/0.29 | 24.35/0.39 |
| 65 | Noisy | 11.81/0.16 | 11.92/0.14 | 11.91/0.18 | 11.92/0.22 | 11.91/0.09 | 11.79/0.22 | 11.95/0.18 | 11.95/0.22 | 11.77/0.17 | 11.77/0.17 | 12.10/0.11 |
| | NL-means | 21.53/0.44 | 21.99/0.39 | 20.95/0.38 | 20.48/0.45 | 22.51/0.32 | 19.75/0.38 | 21.78/0.42 | 20.93/0.43 | 20.48/0.30 | 20.48/0.30 | 22.14/0.31 |
| | K-SVD | 21.64/0.40 | 22.96/0.37 | 21.35/0.33 | 20.46/0.35 | 23.33/0.24 | 19.52/0.30 | 21.93/0.39 | 20.57/0.32 | 20.93/0.19 | 20.93/0.19 | 22.81/0.26 |
| | BM3D | 22.40/0.47 | 23.29/0.44 | 21.95/0.35 | 21.28/0.46 | 24.24/0.36 | 20.47/0.40 | 22.66/0.43 | 21.27/0.38 | 21.06/0.19 | 21.06/0.19 | 23.80/0.33 |
| 80 | Noisy | 10.15/0.12 | 10.15/0.11 | 10.09/0.14 | 10.04/0.17 | 10.13/0.07 | 10.07/0.17 | 10.04/0.12 | 09.96/0.15 | 10.10/0.12 | 10.10/0.12 | 10.13/0.08 |
| | NL-means | 20.49/0.39 | 20.77/0.33 | 20.23/0.34 | 19.44/0.41 | 21.27/0.27 | 19.30/0.38 | 20.13/0.35 | 19.44/0.33 | 19.78/0.27 | 19.78/0.27 | 21.27/0.27 |
| | K-SVD | 21.26/0.36 | 22.18/0.33 | 20.58/0.27 | 19.90/0.32 | 22.97/0.23 | 19.50/0.26 | 20.58/0.32 | 19.72/0.24 | 20.79/0.15 | 20.79/0.15 | 22.58/0.25 |
| | BM3D | 21.60/0.41 | 22.63/0.38 | 21.71/0.34 | 20.66/0.41 | 23.42/0.33 | 20.20/0.41 | 21.85/0.39 | 20.23/0.29 | 20.87/0.17 | 20.87/0.17 | 23.65/0.31 |
| 95 | Noisy | 08.76/0.10 | 08.63/0.08 | 08.55/0.11 | 08.56/0.14 | 08.60/0.05 | 08.51/0.13 | 08.54/0.10 | 08.47/0.11 | 08.50/0.09 | 08.50/0.09 | 08.55/0.05 |
| | NL-means | 19.95/0.37 | 19.45/0.26 | 18.54/0.27 | 18.90/0.38 | 20.36/0.27 | 18.49/0.33 | 19.39/0.33 | 18.68/0.32 | 18.85/0.23 | 18.85/0.23 | 19.92/0.24 |
| | K-SVD | 20.34/0.32 | 21.21/0.29 | 19.80/0.22 | 19.27/0.28 | 23.09/0.25 | 18.91/0.23 | 20.09/0.28 | 19.43/0.25 | 20.40/0.16 | 20.40/0.16 | 22.01/0.23 |
| | BM3D | 21.60/0.40 | 21.40/0.32 | 20.63/0.30 | 20.19/0.36 | 23.35/0.34 | 19.69/0.33 | 21.17/0.34 | 20.14/0.30 | 20.52/0.11 | 20.52/0.11 | 23.13/0.29 |
| 100 | Noisy | 07.29/0.07 | 07.28/0.07 | 07.29/0.09 | 07.35/0.11 | 07.22/0.03 | 07.36/0.10 | 07.39/0.07 | 07.27/0.08 | 07.17/0.07 | 07.17/0.07 | 07.43/0.04 |
| | NL-means | 18.65/0.32 | 18.98/0.27 | 18.02/0.27 | 17.67/0.31 | 18.95/0.17 | 17.65/0.28 | 18.33/0.27 | 17.86/0.28 | 17.99/0.17 | 17.99/0.17 | 19.11/0.18 |
| | K-SVD | 19.85/0.26 | 20.87/0.30 | 19.91/0.25 | 18.63/0.22 | 21.94/0.18 | 18.42/0.15 | 19.45/0.21 | 18.82/0.19 | 20.48/0.12 | 20.48/0.12 | 22.16/0.21 |
| | BM3D | 20.80/0.35 | 22.02/0.37 | 20.44/0.31 | 19.10/0.27 | 22.26/0.18 | 19.26/0.23 | 20.19/0.28 | 19.77/0.26 | 20.15/0.09 | 20.15/0.09 | 22.48/0.21 |

Table 5: Denoising using NL-means, K-SVD and BM3D algorithms when applied on different *128x128* images from *Natural Images Dataset* using multiple noise levels

| Noise | Algorithm | Bee | Bird | Boat | Bridge | Buildings | Cart | House | Owl | Terrain | Tomb | Water |
|---|---|---|---|---|---|---|---|---|---|---|---|---|
| 5 | Noisy | 34.16/0.81 | 34.14/0.76 | 34.25/0.86 | 34.09/0.96 | 34.19/0.83 | 34.25/0.93 | 34.10/0.90 | 34.14/0.95 | 34.05/0.97 | 34.19/0.71 | 34.14/0.86 |
| | NL-means | 34.24/0.79 | 33.80/0.77 | 31.66/0.89 | 29.14/0.89 | 34.68/0.84 | 30.20/0.90 | 31.13/0.84 | 29.33/0.87 | 25.58/0.83 | 33.36/0.64 | 29.86/0.77 |
| | K-SVD | 36.80/0.85 | 37.22/0.81 | 36.33/0.92 | 34.82/0.96 | 36.84/0.88 | 35.46/0.95 | 35.57/0.91 | 34.88/0.95 | 34.24/0.97 | 36.76/0.72 | 35.40/0.87 |
| | BM3D | 37.15/0.83 | 37.41/0.79 | 36.57/0.93 | 34.89/0.96 | 36.98/0.88 | 35.69/0.94 | 35.71/0.89 | 34.91/0.95 | 34.32/0.97 | 36.83/0.67 | 35.51/0.85 |
| 20 | Noisy | 22.15/0.40 | 22.05/0.37 | 22.21/0.50 | 22.17/0.66 | 22.14/0.37 | 22.09/0.63 | 21.97/0.56 | 22.15/0.64 | 22.20/0.71 | 22.03/0.33 | 22.14/0.50 |
| | NL-means | 29.58/0.61 | 29.29/0.59 | 27.46/0.65 | 25.60/0.69 | 29.14/0.57 | 26.30/0.74 | 27.10/0.70 | 25.79/0.67 | 23.17/0.61 | 28.65/0.43 | 26.12/0.52 |
| | K-SVD | 29.96/0.61 | 29.90/0.61 | 28.10/0.66 | 26.25/0.74 | 29.74/0.58 | 26.76/0.77 | 27.66/0.72 | 26.32/0.71 | 24.68/0.71 | 29.81/0.47 | 27.23/0.60 |
| | BM3D | 30.29/0.62 | 30.11/0.62 | 28.16/0.69 | 26.22/0.73 | 30.42/0.65 | 27.09/0.79 | 27.91/0.73 | 26.38/0.70 | 24.48/0.67 | 29.87/0.45 | 27.02/0.58 |
| 35 | Noisy | 17.21/0.24 | 17.22/0.22 | 17.27/0.33 | 17.26/0.44 | 17.18/0.19 | 17.23/0.43 | 16.06/0.35 | 17.21/0.41 | 17.28/0.48 | 17.21/0.19 | 17.42/0.30 |
| | NL-means | 26.41/0.49 | 26.21/0.45 | 24.70/0.52 | 23.45/0.57 | 26.29/0.42 | 23.39/0.59 | 24.34/0.54 | 23.32/0.53 | 21.22/0.45 | 25.85/0.32 | 24.19/0.38 |
| | K-SVD | 27.16/0.50 | 26.93/0.46 | 25.27/0.53 | 23.85/0.58 | 26.93/0.40 | 23.70/0.60 | 24.52/0.55 | 23.57/0.52 | 21.90/0.49 | 27.13/0.34 | 24.60/0.38 |
| | BM3D | 27.85/0.53 | 27.33/0.51 | 25.47/0.56 | 23.94/0.60 | 27.87/0.51 | 24.17/0.65 | 24.97/0.58 | 23.71/0.54 | 21.43/0.44 | 27.55/0.36 | 24.88/0.42 |
| 50 | Noisy | 14.15/0.16 | 14.14/0.15 | 14.07/0.22 | 14.16/0.29 | 14.06/0.12 | 14.12/0.30 | 14.15/0.25 | 14.10/0.28 | 14.18/0.34 | 14.14/0.12 | 14.30/0.19 |
| | NL-means | 24.22/0.39 | 24.16/0.37 | 22.72/0.41 | 21.84/0.49 | 24.42/0.33 | 21.77/0.49 | 22.60/0.45 | 22.04/0.45 | 20.14/0.37 | 24.08/0.26 | 22.84/0.30 |
| | K-SVD | 24.97/0.41 | 25.16/0.39 | 23.45/0.42 | 22.07/0.45 | 25.53/0.32 | 22.08/0.49 | 22.64/0.43 | 22.05/0.39 | 20.42/0.34 | 25.46/0.27 | 23.14/0.25 |
| | BM3D | 25.90/0.46 | 25.58/0.42 | 23.73/0.44 | 22.56/0.50 | 26.53/0.42 | 22.47/0.55 | 23.52/0.49 | 22.47/0.44 | 20.07/0.29 | 25.99/0.28 | 23.65/0.29 |
| 65 | Noisy | 11.77/0.11 | 11.83/0.10 | 11.93/0.16 | 11.96/0.21 | 11.83/0.07 | 11.88/0.21 | 11.95/0.18 | 11.89/0.20 | 11.83/0.25 | 11.96/0.09 | 11.85/0.12 |
| | NL-means | 22.61/0.34 | 22.71/0.31 | 21.45/0.36 | 20.66/0.41 | 23.00/0.27 | 20.51/0.42 | 21.78/0.42 | 20.95/0.41 | 19.31/0.32 | 22.77/0.22 | 21.51/0.25 |
| | K-SVD | 23.44/0.34 | 24.22/0.33 | 22.33/0.36 | 20.80/0.34 | 24.67/0.27 | 20.66/0.37 | 21.93/0.39 | 20.95/0.30 | 19.55/0.26 | 24.42/0.24 | 22.49/0.20 |
| | BM3D | 25.02/0.42 | 24.87/0.40 | 22.82/0.40 | 21.61/0.42 | 25.62/0.34 | 21.35/0.46 | 22.66/0.43 | 21.64/0.38 | 19.47/0.23 | 25.35/0.27 | 22.96/0.25 |
| 80 | Noisy | 10.14/0.08 | 10.08/0.08 | 10.11/0.12 | 10.14/0.15 | 10.14/0.05 | 10.09/0.15 | 10.04/0.12 | 10.01/0.14 | 10.03/0.18 | 10.00/0.06 | 10.00/0.09 |
| | NL-means | 21.43/0.29 | 21.18/0.27 | 20.26/0.31 | 19.72/0.35 | 21.68/0.23 | 19.39/0.34 | 20.13/0.35 | 19.88/0.34 | 18.41/0.25 | 21.11/0.18 | 20.48/0.21 |
| | K-SVD | 22.80/0.31 | 23.08/0.29 | 21.17/0.27 | 20.26/0.29 | 24.12/0.25 | 19.59/0.27 | 20.58/0.32 | 20.29/0.25 | 18.93/0.18 | 23.16/0.18 | 21.94/0.17 |
| | BM3D | 24.21/0.38 | 23.73/0.34 | 22.06/0.34 | 20.97/0.35 | 24.82/0.29 | 20.49/0.38 | 21.85/0.39 | 21.02/0.31 | 18.89/0.16 | 24.28/0.22 | 22.62/0.22 |
| 95 | Noisy | 08.56/0.06 | 08.52/0.06 | 08.59/0.09 | 08.55/0.12 | 08.61/0.04 | 08.55/0.12 | 08.54/0.10 | 08.56/0.11 | 08.64/0.14 | 08.61/0.04 | 08.52/0.06 |
| | NL-means | 20.01/0.25 | 19.80/0.22 | 19.23/0.26 | 18.76/0.33 | 20.38/0.18 | 18.69/0.31 | 19.39/0.33 | 19.20/0.31 | 17.83/0.25 | 20.08/0.15 | 19.51/0.19 |
| | K-SVD | 21.80/0.26 | 22.28/0.28 | 20.35/0.24 | 19.76/0.28 | 23.16/0.20 | 19.19/0.24 | 20.09/0.28 | 20.05/0.22 | 18.70/0.15 | 23.24/0.18 | 21.58/0.16 |
| | BM3D | 23.20/0.32 | 22.93/0.31 | 21.43/0.30 | 20.53/0.36 | 23.85/0.21 | 20.17/0.35 | 21.17/0.34 | 20.71/0.28 | 18.78/0.15 | 23.86/0.20 | 22.25/0.19 |
| 100 | Noisy | 07.34/0.04 | 07.30/0.05 | 07.33/0.07 | 07.39/0.09 | 07.35/0.03 | 07.30/0.09 | 07.39/0.07 | 07.35/0.08 | 07.31/0.12 | 07.25/0.03 | 07.30/0.05 |
| | NL-means | 19.18/0.21 | 19.01/0.19 | 18.55/0.23 | 18.07/0.28 | 19.20/0.15 | 17.83/0.28 | 18.33/0.27 | 18.28/0.27 | 17.28/0.23 | 19.06/0.13 | 18.58/0.17 |
| | K-SVD | 21.75/0.25 | 22.04/0.24 | 20.16/0.24 | 19.45/0.23 | 23.11/0.18 | 18.80/0.21 | 19.45/0.21 | 19.56/0.19 | 18.46/0.14 | 22.53/0.16 | 21.25/0.13 |
| | BM3D | 22.81/0.29 | 22.56/0.26 | 21.32/0.28 | 19.96/0.29 | 23.63/0.20 | 19.52/0.31 | 20.19/0.28 | 20.16/0.24 | 18.53/0.12 | 23.66/0.19 | 21.79/0.16 |

Table 6: Denoising using NL-means, K-SVD and BM3D algorithms when applied on different *256x256* images from *Natural Images Dataset* using multiple noise levels

| Noise | Algorithm | Bee | Bird | Boat | Bridge | Buildings | Cart | House | Owl | Terrain | Tomb | Water |
|---|---|---|---|---|---|---|---|---|---|---|---|---|
| 5 | Noisy | 34.13/0.94 | 34.17/0.95 | 34.13/0.83 | 34.14/0.98 | 34.13/0.95 | 34.12/0.97 | 34.16/0.98 | 34.13/0.98 | 34.16/0.99 | 34.16/0.97 | 34.19/0.96 |
| | NL-means | 28.63/0.80 | 28.23/0.86 | 30.86/0.83 | 24.68/0.85 | 28.10/0.83 | 25.60/0.86 | 24.16/0.84 | 25.27/0.85 | 20.59/0.84 | 27.36/0.86 | 24.08/0.82 |
| | K-SVD | 34.59/0.94 | 34.64/0.96 | 36.13/0.87 | 34.19/0.98 | 34.56/0.95 | 34.30/0.97 | 34.26/0.98 | 34.20/0.98 | 34.09/0.99 | 34.49/0.97 | 34.37/0.96 |
| | BM3D | 34.79/0.94 | 34.86/0.96 | 36.33/0.87 | 34.33/0.98 | 34.69/0.95 | 34.83/0.98 | 34.53/0.98 | 34.39/0.98 | 34.25/0.99 | 34.73/0.97 | 34.59/0.96 |
| 20 | Noisy | 22.09/0.60 | 22.10/0.66 | 22.13/0.48 | 22.08/0.79 | 22.16/0.64 | 22.13/0.76 | 22.12/0.77 | 22.10/0.78 | 22.14/0.90 | 22.12/0.69 | 22.11/0.74 |
| | NL-means | 24.58/0.47 | 24.15/0.53 | 27.26/0.60 | 21.92/0.61 | 24.02/0.48 | 22.80/0.64 | 21.51/0.56 | 22.44/0.62 | 18.99/0.68 | 23.42/0.51 | 21.43/0.53 |
| | K-SVD | 25.62/0.58 | 25.36/0.65 | 28.03/0.63 | 23.92/0.78 | 25.40/0.62 | 24.35/0.76 | 23.93/0.75 | 24.06/0.77 | 22.86/0.89 | 24.91/0.67 | 24.02/0.72 |
| | BM3D | 25.38/0.51 | 25.23/0.61 | 28.21/0.65 | 23.77/0.76 | 25.00/0.55 | 24.30/0.73 | 23.68/0.71 | 23.95/0.75 | 22.85/0.88 | 24.68/0.62 | 23.85/0.68 |
| 35 | Noisy | 17.23/0.37 | 17.23/0.41 | 17.29/0.31 | 17.26/0.57 | 17.28/0.40 | 17.25/0.54 | 17.26/0.55 | 17.28/0.55 | 17.25/0.74 | 17.25/0.44 | 17.26/0.54 |
| | NL-means | 23.11/0.35 | 22.50/0.38 | 24.49/0.46 | 20.37/0.47 | 22.46/0.33 | 20.99/0.50 | 20.03/0.41 | 20.87/0.47 | 17.38/0.51 | 21.85/0.34 | 19.88/0.37 |
| | K-SVD | 23.53/0.34 | 22.87/0.37 | 25.39/0.48 | 21.06/0.54 | 23.00/0.33 | 21.57/0.53 | 21.09/0.49 | 21.31/0.51 | 19.18/0.70 | 22.42/0.35 | 21.03/0.47 |
| | BM3D | 23.49/0.30 | 22.92/0.35 | 25.57/0.52 | 20.78/0.49 | 22.92/0.31 | 21.47/0.51 | 20.85/0.45 | 21.09/0.47 | 18.75/0.65 | 22.34/0.32 | 20.61/0.40 |
| 50 | Noisy | 14.12/0.24 | 14.13/0.27 | 14.20/0.21 | 14.13/0.41 | 14.13/0.26 | 14.19/0.39 | 14.18/0.39 | 14.11/0.38 | 14.14/0.60 | 14.16/0.29 | 14.17/0.39 |
| | NL-means | 22.02/0.29 | 21.42/0.31 | 22.74/0.37 | 19.44/0.39 | 21.45/0.27 | 19.75/0.41 | 19.06/0.32 | 19.97/0.40 | 16.43/0.41 | 20.88/0.27 | 18.98/0.29 |
| | K-SVD | 22.65/0.24 | 21.95/0.25 | 23.79/0.39 | 19.80/0.38 | 22.08/0.22 | 20.17/0.39 | 19.66/0.32 | 20.08/0.36 | 17.34/0.52 | 21.47/0.22 | 19.62/0.30 |
| | BM3D | 22.75/0.22 | 22.01/0.24 | 23.87/0.42 | 19.64/0.35 | 22.15/0.21 | 20.12/0.38 | 19.35/0.28 | 20.04/0.34 | 16.76/0.41 | 21.45/0.20 | 19.22/0.23 |
| 65 | Noisy | 11.87/0.16 | 11.87/0.18 | 11.88/0.16 | 11.84/0.30 | 11.84/0.17 | 11.83/0.28 | 11.88/0.28 | 11.88/0.27 | 11.90/0.48 | 11.88/0.20 | 11.92/0.29 |
| | NL-means | 21.11/0.26 | 20.42/0.26 | 21.26/0.31 | 18.65/0.34 | 20.52/0.22 | 18.87/0.35 | 18.39/0.28 | 19.26/0.35 | 15.78/0.34 | 20.02/0.23 | 18.31/0.25 |
| | K-SVD | 22.04/0.20 | 21.31/0.19 | 22.49/0.32 | 19.02/0.29 | 21.46/0.16 | 19.31/0.32 | 18.97/0.23 | 19.36/0.26 | 16.27/0.37 | 20.82/0.16 | 18.77/0.20 |
| | BM3D | 22.39/0.20 | 21.44/0.19 | 22.75/0.35 | 19.05/0.28 | 21.65/0.16 | 19.46/0.32 | 18.82/0.21 | 19.52/0.27 | 15.79/0.27 | 20.98/0.15 | 18.61/0.16 |
| 80 | Noisy | 10.06/0.12 | 10.09/0.14 | 10.04/0.11 | 10.09/0.23 | 10.10/0.12 | 10.07/0.21 | 10.03/0.21 | 10.06/0.20 | 10.08/0.39 | 10.05/0.14 | 10.06/0.21 |
| | NL-means | 20.10/0.23 | 19.56/0.23 | 20.13/0.26 | 18.04/0.31 | 19.71/0.20 | 18.15/0.31 | 17.78/0.25 | 18.57/0.32 | 15.29/0.31 | 19.24/0.19 | 17.68/0.22 |
| | K-SVD | 21.41/0.16 | 20.87/0.17 | 21.49/0.27 | 18.46/0.23 | 21.04/0.12 | 18.59/0.25 | 18.46/0.19 | 18.89/0.21 | 15.60/0.28 | 20.41/0.12 | 18.33/0.16 |
| | BM3D | 22.01/0.18 | 21.18/0.17 | 22.00/0.30 | 18.73/0.25 | 21.25/0.13 | 18.97/0.28 | 18.44/0.18 | 19.15/0.24 | 15.27/0.20 | 20.70/0.12 | 18.31/0.13 |
| 95 | Noisy | 08.61/0.09 | 08.60/0.10 | 08.58/0.09 | 08.59/0.17 | 08.61/0.09 | 08.58/0.16 | 08.53/0.16 | 08.61/0.15 | 08.60/0.32 | 08.64/0.10 | 08.57/0.17 |
| | NL-means | 19.17/0.20 | 18.71/0.19 | 19.24/0.23 | 17.35/0.28 | 18.84/0.17 | 17.41/0.27 | 17.10/0.21 | 17.85/0.28 | 14.87/0.28 | 18.46/0.17 | 17.09/0.19 |
| | K-SVD | 21.04/0.15 | 20.48/0.14 | 20.84/0.23 | 18.08/0.19 | 20.81/0.11 | 17.99/0.20 | 18.03/0.15 | 18.52/0.18 | 15.14/0.21 | 19.98/0.09 | 18.03/0.12 |
| | BM3D | 21.69/0.17 | 20.78/0.15 | 21.51/0.27 | 18.37/0.22 | 21.07/0.12 | 18.52/0.24 | 18.11/0.15 | 18.81/0.21 | 14.96/0.16 | 20.32/0.10 | 18.08/0.11 |
| 100 | Noisy | 07.27/0.07 | 07.28/0.08 | 07.29/0.07 | 07.29/0.14 | 07.30/0.07 | 07.27/0.13 | 07.32/0.13 | 07.31/0.11 | 07.30/0.26 | 07.32/0.08 | 07.29/0.13 |
| | NL-means | 18.30/0.18 | 17.85/0.17 | 18.26/0.20 | 16.78/0.26 | 18.06/0.15 | 16.73/0.24 | 16.61/0.19 | 17.17/0.26 | 14.45/0.25 | 17.83/0.15 | 16.56/0.17 |
| | K-SVD | 20.76/0.14 | 20.04/0.14 | 20.39/0.21 | 17.74/0.17 | 20.49/0.10 | 17.61/0.17 | 17.81/0.13 | 18.24/0.15 | 14.78/0.16 | 19.80/0.09 | 17.74/0.10 |
| | BM3D | 21.34/0.16 | 20.47/0.14 | 21.00/0.23 | 18.07/0.20 | 20.80/0.10 | 18.18/0.22 | 17.97/0.13 | 18.51/0.18 | 14.71/0.13 | 20.25/0.09 | 17.87/0.08 |

Table 7: Denoising using NL-means, K-SVD and BM3D algorithms when applied on different *64x64* images from *Texture Test Dataset* using multiple noise levels

| Noise | Algorithm | 1 | 2 | 3 | 4 | 5 | 6 | 7 | 8 | 9 | 10 |
|---|---|---|---|---|---|---|---|---|---|---|---|
| 5 | Noisy | 34.21/0.99 | 34.17/0.97 | 34.09/0.97 | 34.17/0.99 | 34.18/0.99 | 34.29/0.99 | 34.10/0.90 | 34.08/0.94 | 34.29/0.99 | 34.36/0.98 |
| | NL-means | 20.74/0.84 | 27.75/0.85 | 27.08/0.80 | 23.85/0.81 | 20.89/0.85 | 23.11/0.83 | 30.96/0.69 | 30.11/0.82 | 23.11/0.83 | 26.22/0.88 |
| | K-SVD | 34.24/0.99 | 34.66/0.97 | 34.28/0.97 | 34.16/0.98 | 34.17/0.99 | 34.29/0.99 | 34.95/0.90 | 34.75/0.94 | 34.29/0.99 | 34.48/0.98 |
| | BM3D | 34.25/0.99 | 34.70/0.97 | 34.38/0.97 | 34.21/0.99 | 34.18/0.99 | 34.35/0.99 | 35.02/0.90 | 34.77/0.94 | 34.35/0.99 | 34.52/0.98 |
| 20 | Noisy | 22.13/0.91 | 22.04/0.68 | 22.21/0.68 | 22.18/0.81 | 22.03/0.91 | 22.12/0.85 | 22.20/0.40 | 22.07/0.52 | 22.12/0.85 | 21.85/0.77 |
| | NL-means | 20.09/0.79 | 24.19/0.57 | 23.71/0.47 | 21.17/0.54 | 19.89/0.78 | 21.20/0.70 | 27.25/0.28 | 25.67/0.40 | 21.20/0.70 | 22.81/0.66 |
| | K-SVD | 23.32/0.92 | 25.35/0.71 | 25.15/0.68 | 23.75/0.81 | 23.22/0.91 | 23.59/0.86 | 27.68/0.34 | 26.23/0.50 | 23.59/0.86 | 24.24/0.80 |
| | BM3D | 23.18/0.92 | 25.08/0.68 | 24.67/0.60 | 23.52/0.79 | 23.10/0.91 | 23.45/0.85 | 27.63/0.23 | 25.88/0.41 | 23.45/0.85 | 24.17/0.79 |
| 35 | Noisy | 17.09/0.76 | 17.18/0.42 | 17.26/0.40 | 17.18/0.58 | 17.27/0.77 | 17.40/0.66 | 17.25/0.18 | 17.25/0.27 | 17.40/0.66 | 17.28/0.55 |
| | NL-means | 18.06/0.62 | 22.53/0.42 | 22.33/0.32 | 19.66/0.34 | 18.14/0.64 | 19.26/0.49 | 25.72/0.20 | 24.26/0.28 | 19.26/0.49 | 20.82/0.45 |
| | K-SVD | 19.89/0.80 | 22.44/0.42 | 22.48/0.34 | 20.69/0.53 | 19.86/0.80 | 20.42/0.66 | 26.53/0.11 | 24.59/0.23 | 20.42/0.66 | 21.50/0.56 |
| | BM3D | 19.23/0.76 | 22.63/0.41 | 22.39/0.25 | 20.16/0.41 | 19.32/0.76 | 20.01/0.61 | 26.56/0.08 | 24.59/0.19 | 20.01/0.61 | 21.42/0.55 |
| 50 | Noisy | 13.93/0.61 | 14.26/0.28 | 14.13/0.25 | 14.10/0.41 | 14.02/0.61 | 14.19/0.47 | 14.24/0.10 | 14.11/0.17 | 14.19/0.47 | 14.21/0.40 |
| | NL-means | 16.61/0.46 | 21.76/0.37 | 21.56/0.26 | 18.97/0.28 | 16.58/0.45 | 18.22/0.37 | 24.06/0.14 | 23.22/0.27 | 18.22/0.37 | 19.76/0.36 |
| | K-SVD | 17.75/0.65 | 21.68/0.26 | 21.73/0.19 | 19.29/0.32 | 17.69/0.63 | 18.57/0.43 | 26.11/0.09 | 23.88/0.18 | 18.57/0.43 | 20.03/0.37 |
| | BM3D | 17.60/0.60 | 21.84/0.25 | 21.71/0.11 | 19.03/0.17 | 17.35/0.56 | 18.36/0.35 | 26.28/0.04 | 24.28/0.14 | 18.36/0.35 | 20.13/0.36 |
| 65 | Noisy | 11.90/0.50 | 11.87/0.18 | 12.10/0.18 | 11.77/0.30 | 11.79/0.47 | 11.93/0.36 | 11.93/0.06 | 11.92/0.11 | 11.93/0.36 | 12.02/0.28 |
| | NL-means | 15.93/0.38 | 20.49/0.29 | 20.72/0.23 | 18.37/0.25 | 15.70/0.33 | 17.65/0.32 | 22.95/0.13 | 21.89/0.18 | 17.65/0.32 | 19.03/0.29 |
| | K-SVD | 16.67/0.51 | 21.01/0.16 | 21.41/0.10 | 18.79/0.23 | 16.13/0.42 | 17.78/0.31 | 25.59/0.07 | 23.51/0.11 | 17.78/0.31 | 19.35/0.20 |
| | BM3D | 16.34/0.42 | 21.09/0.16 | 21.43/0.08 | 18.73/0.13 | 15.86/0.33 | 17.64/0.23 | 26.18/0.03 | 23.83/0.10 | 17.64/0.23 | 19.39/0.24 |
| 80 | Noisy | 10.19/0.41 | 10.18/0.12 | 10.12/0.13 | 10.11/0.22 | 10.10/0.39 | 10.14/0.28 | 10.08/0.04 | 10.19/0.07 | 10.14/0.28 | 10.01/0.18 |
| | NL-means | 15.33/0.31 | 19.83/0.25 | 19.83/0.21 | 17.78/0.21 | 15.25/0.30 | 17.32/0.31 | 21.27/0.10 | 20.69/0.14 | 17.32/0.31 | 18.12/0.20 |
| | K-SVD | 15.63/0.35 | 20.80/0.10 | 21.33/0.09 | 18.44/0.12 | 15.41/0.31 | 17.22/0.22 | 25.13/0.06 | 23.17/0.09 | 17.22/0.22 | 18.79/0.14 |
| | BM3D | 15.50/0.27 | 20.68/0.10 | 21.35/0.07 | 18.51/0.07 | 15.24/0.20 | 17.50/0.20 | 25.94/0.04 | 23.33/0.08 | 17.50/0.20 | 18.74/0.10 |
| 95 | Noisy | 08.44/0.31 | 08.65/0.09 | 08.61/0.08 | 08.61/0.15 | 08.58/0.33 | 08.41/0.21 | 08.63/0.03 | 08.59/0.05 | 08.41/0.21 | 08.90/0.16 |
| | NL-means | 14.90/0.30 | 18.88/0.23 | 18.60/0.15 | 17.33/0.17 | 14.84/0.31 | 16.60/0.27 | 19.87/0.07 | 19.47/0.14 | 16.60/0.27 | 17.76/0.18 |
| | K-SVD | 15.21/0.29 | 20.39/0.10 | 20.85/0.07 | 18.31/0.07 | 15.29/0.30 | 16.78/0.18 | 24.36/0.04 | 22.73/0.08 | 16.78/0.18 | 18.66/0.08 |
| | BM3D | 15.33/0.24 | 20.62/0.08 | 20.87/0.06 | 18.42/0.05 | 15.31/0.26 | 17.11/0.16 | 25.24/0.03 | 22.91/0.05 | 17.11/0.16 | 18.53/0.05 |
| 100 | Noisy | 07.33/0.27 | 07.41/0.06 | 07.38/0.07 | 07.23/0.12 | 07.24/0.26 | 07.10/0.16 | 07.32/0.04 | 07.35/0.05 | 07.10/0.16 | 07.30/0.11 |
| | NL-means | 14.55/0.27 | 17.93/0.19 | 18.04/0.13 | 16.77/0.17 | 14.60/0.26 | 15.95/0.24 | 19.25/0.10 | 18.79/0.12 | 15.95/0.24 | 16.99/0.19 |
| | K-SVD | 14.96/0.21 | 20.22/0.08 | 20.71/0.05 | 18.15/0.08 | 14.82/0.21 | 16.69/0.18 | 23.68/0.05 | 22.57/0.08 | 16.69/0.18 | 18.55/0.08 |
| | BM3D | 14.96/0.17 | 20.20/0.07 | 21.03/0.05 | 18.39/0.05 | 14.97/0.16 | 16.95/0.13 | 25.11/0.06 | 23.01/0.06 | 16.95/0.13 | 18.70/0.09 |

Table 8: Denoising using NL-means, K-SVD and BM3D algorithms when applied on different *128x128* images from *Texture Test Dataset* using multiple noise levels

| Noise | Algorithm | 1 | 2 | 3 | 4 | 5 | 6 | 7 | 8 | 9 | 10 |
|---|---|---|---|---|---|---|---|---|---|---|---|
| 5 | Noisy | 34.20/0.99 | 34.20/0.98 | 34.27/0.99 | 34.12/0.99 | 34.10/0.99 | 34.10/0.99 | 34.13/0.99 | 34.09/0.96 | 34.06/0.97 | 34.18/0.99 |
| | NL-means | 22.29/0.82 | 27.00/0.87 | 24.27/0.83 | 19.93/0.83 | 21.35/0.90 | 21.33/0.86 | 20.37/0.83 | 28.09/0.79 | 27.98/0.88 | 25.43/0.90 |
| | K-SVD | 34.19/0.99 | 34.52/0.98 | 34.29/0.99 | 34.09/0.99 | 34.16/0.99 | 34.14/0.99 | 34.11/0.99 | 34.43/0.96 | 34.50/0.97 | 34.40/0.99 |
| | BM3D | 34.23/0.99 | 34.59/0.98 | 34.36/0.99 | 34.14/0.99 | 34.19/0.99 | 34.16/0.99 | 34.16/0.99 | 34.52/0.96 | 34.47/0.97 | 34.44/0.99 |
| 20 | Noisy | 22.13/0.86 | 22.05/0.74 | 22.17/0.81 | 22.14/0.92 | 22.04/0.92 | 22.02/0.90 | 22.10/0.91 | 22.17/0.61 | 22.10/0.71 | 22.08/0.83 |
| | NL-means | 20.46/0.66 | 23.81/0.67 | 21.78/0.62 | 18.82/0.74 | 20.52/0.86 | 20.41/0.81 | 19.31/0.75 | 24.36/0.40 | 24.18/0.63 | 23.04/0.79 |
| | K-SVD | 23.35/0.86 | 25.30/0.80 | 23.90/0.81 | 22.85/0.92 | 23.29/0.94 | 23.43/0.92 | 22.94/0.91 | 25.62/0.58 | 25.22/0.75 | 24.34/0.86 |
| | BM3D | 23.22/0.86 | 25.14/0.79 | 23.73/0.80 | 22.64/0.92 | 23.23/0.93 | 23.23/0.91 | 22.81/0.91 | 24.93/0.47 | 24.89/0.71 | 24.37/0.86 |
| 35 | Noisy | 17.30/0.68 | 17.33/0.50 | 17.25/0.59 | 17.25/0.79 | 17.23/0.81 | 17.25/0.76 | 17.21/0.77 | 17.31/0.36 | 17.23/0.46 | 17.43/0.64 |
| | NL-means | 18.66/0.45 | 21.71/0.47 | 20.10/0.43 | 17.01/0.55 | 18.60/0.76 | 18.14/0.63 | 17.51/0.58 | 23.08/0.28 | 22.05/0.43 | 20.56/0.61 |
| | K-SVD | 20.07/0.66 | 22.41/0.54 | 20.99/0.57 | 19.09/0.78 | 19.91/0.85 | 20.01/0.80 | 19.28/0.77 | 23.51/0.27 | 22.39/0.44 | 21.44/0.70 |
| | BM3D | 19.51/0.60 | 22.42/0.56 | 20.62/0.52 | 18.48/0.74 | 19.65/0.84 | 19.37/0.77 | 18.75/0.73 | 23.35/0.21 | 22.11/0.39 | 21.32/0.70 |
| 50 | Noisy | 14.17/0.51 | 14.09/0.33 | 14.20/0.42 | 14.23/0.65 | 14.16/0.68 | 14.10/0.62 | 14.12/0.63 | 14.06/0.21 | 14.16/0.31 | 14.05/0.46 |
| | NL-means | 17.83/0.35 | 20.70/0.38 | 19.19/0.34 | 15.97/0.42 | 17.04/0.65 | 16.78/0.49 | 16.39/0.45 | 22.09/0.23 | 20.91/0.32 | 19.05/0.48 |
| | K-SVD | 18.40/0.44 | 20.88/0.35 | 19.40/0.33 | 17.22/0.62 | 18.01/0.75 | 18.12/0.67 | 17.49/0.62 | 22.64/0.16 | 21.13/0.25 | 19.56/0.53 |
| | BM3D | 17.87/0.31 | 20.99/0.36 | 19.26/0.29 | 16.62/0.51 | 17.90/0.74 | 17.87/0.63 | 17.04/0.54 | 22.80/0.11 | 21.09/0.21 | 19.44/0.52 |
| 65 | Noisy | 11.92/0.39 | 11.75/0.23 | 11.81/0.29 | 11.82/0.52 | 11.90/0.56 | 11.98/0.50 | 11.88/0.50 | 11.85/0.14 | 11.84/0.21 | 11.88/0.35 |
| | NL-means | 17.22/0.29 | 19.85/0.33 | 18.51/0.29 | 15.31/0.35 | 16.07/0.56 | 16.03/0.41 | 15.71/0.36 | 21.13/0.18 | 20.04/0.27 | 18.30/0.42 |
| | K-SVD | 17.43/0.27 | 20.01/0.25 | 18.73/0.23 | 16.02/0.46 | 16.60/0.63 | 16.79/0.50 | 16.21/0.43 | 22.30/0.11 | 20.50/0.18 | 18.37/0.37 |
| | BM3D | 17.20/0.18 | 20.32/0.28 | 18.62/0.16 | 15.51/0.34 | 16.52/0.62 | 16.64/0.48 | 15.82/0.34 | 22.58/0.08 | 20.61/0.14 | 18.66/0.43 |
| 80 | Noisy | 10.07/0.30 | 10.03/0.16 | 10.04/0.22 | 10.22/0.43 | 10.09/0.46 | 10.04/0.39 | 10.08/0.41 | 10.04/0.10 | 10.16/0.15 | 10.12/0.27 |
| | NL-means | 16.74/0.27 | 18.98/0.28 | 17.90/0.26 | 14.89/0.30 | 15.47/0.52 | 15.47/0.35 | 15.32/0.33 | 20.13/0.17 | 19.26/0.23 | 17.48/0.36 |
| | K-SVD | 17.03/0.20 | 19.53/0.17 | 18.30/0.16 | 15.16/0.30 | 15.51/0.51 | 15.76/0.36 | 15.50/0.31 | 21.83/0.10 | 20.19/0.11 | 17.49/0.25 |
| | BM3D | 16.95/0.15 | 19.77/0.19 | 18.38/0.12 | 14.90/0.21 | 15.72/0.54 | 15.74/0.34 | 15.37/0.26 | 22.36/0.07 | 20.27/0.10 | 17.92/0.34 |
| 95 | Noisy | 08.67/0.24 | 08.46/0.13 | 08.58/0.17 | 08.61/0.35 | 08.51/0.37 | 08.54/0.32 | 08.50/0.32 | 08.55/0.07 | 08.59/0.11 | 08.62/0.21 |
| | NL-means | 16.25/0.24 | 18.27/0.27 | 17.44/0.24 | 14.44/0.26 | 14.89/0.47 | 15.10/0.33 | 14.85/0.30 | 19.14/0.14 | 18.38/0.18 | 17.06/0.34 |
| | K-SVD | 16.68/0.14 | 19.25/0.18 | 18.13/0.13 | 14.73/0.22 | 14.70/0.41 | 15.25/0.28 | 15.07/0.24 | 21.51/0.08 | 19.92/0.09 | 17.08/0.19 |
| | BM3D | 16.70/0.10 | 19.61/0.20 | 18.24/0.10 | 14.52/0.13 | 15.07/0.46 | 15.38/0.29 | 14.98/0.18 | 22.06/0.05 | 20.15/0.07 | 17.63/0.27 |
| 100 | Noisy | 07.39/0.19 | 07.23/0.09 | 07.30/0.14 | 07.25/0.28 | 07.28/0.31 | 07.35/0.26 | 07.36/0.27 | 07.30/0.06 | 07.27/0.09 | 07.30/0.17 |
| | NL-means | 15.83/0.23 | 17.43/0.22 | 16.75/0.21 | 14.13/0.25 | 14.44/0.44 | 14.65/0.29 | 14.54/0.27 | 18.37/0.11 | 17.76/0.17 | 16.38/0.30 |
| | K-SVD | 16.50/0.11 | 18.92/0.12 | 17.88/0.12 | 14.49/0.18 | 13.98/0.30 | 14.88/0.21 | 14.76/0.17 | 21.37/0.06 | 19.71/0.09 | 16.80/0.17 |
| | BM3D | 16.61/0.09 | 19.09/0.11 | 18.03/0.07 | 14.41/0.12 | 14.50/0.39 | 15.02/0.22 | 14.71/0.12 | 22.07/0.04 | 20.08/0.07 | 17.29/0.24 |

Table 9: Denoising using NL-means, K-SVD and BM3D algorithms when applied on different *256x256* images from *Texture Test Dataset* using multiple noise levels

| Noise | Algorithm | 1 | 2 | 3 | 4 | 5 | 6 | 7 | 8 | 9 | 10 |
|---|---|---|---|---|---|---|---|---|---|---|---|
| 5 | Noisy | 34.14/1.00 | 34.16/0.99 | 34.16/0.99 | 34.15/1.00 | 34.14/0.99 | 34.18/1.00 | 34.14/1.00 | 34.14/0.98 | 34.13/0.99 | 34.16/0.99 |
| | NL-means | 18.73/0.84 | 22.98/0.83 | 20.38/0.84 | 16.87/0.85 | 20.99/0.91 | 18.64/0.86 | 17.05/0.85 | 25.28/0.83 | 25.27/0.92 | 23.94/0.88 |
| | K-SVD | 34.06/1.00 | 34.16/0.99 | 34.09/0.99 | 33.93/1.00 | 34.12/0.99 | 34.04/1.00 | 33.92/1.00 | 34.26/0.98 | 34.35/0.99 | 34.26/0.99 |
| | BM3D | 34.18/1.00 | 34.23/0.99 | 34.19/0.99 | 34.43/1.00 | 34.40/1.00 | 34.34/1.00 | 34.42/1.00 | 34.33/0.98 | 34.37/0.99 | 34.33/0.99 |
| 20 | Noisy | 22.12/0.94 | 22.16/0.84 | 22.09/0.91 | 22.09/0.96 | 22.16/0.93 | 22.10/0.94 | 22.08/0.96 | 22.13/0.76 | 22.07/0.86 | 22.12/0.85 |
| | NL-means | 18.13/0.80 | 20.75/0.65 | 19.20/0.75 | 16.53/0.82 | 20.18/0.88 | 17.87/0.81 | 16.68/0.82 | 22.43/0.57 | 23.10/0.84 | 22.13/0.78 |
| | K-SVD | 22.68/0.94 | 23.43/0.84 | 22.83/0.91 | 22.41/0.96 | 23.14/0.94 | 22.61/0.94 | 22.41/0.96 | 24.29/0.76 | 24.28/0.89 | 23.93/0.87 |
| | BM3D | 22.62/0.94 | 23.33/0.83 | 22.74/0.91 | 22.62/0.96 | 23.33/0.94 | 22.69/0.94 | 22.61/0.96 | 24.07/0.74 | 24.28/0.89 | 23.95/0.87 |
| 35 | Noisy | 17.25/0.84 | 17.25/0.64 | 17.25/0.78 | 17.27/0.89 | 17.22/0.82 | 17.22/0.84 | 17.27/0.89 | 17.27/0.53 | 17.22/0.67 | 17.22/0.67 |
| | NL-means | 16.47/0.66 | 19.21/0.49 | 17.32/0.57 | 15.36/0.74 | 18.32/0.79 | 16.38/0.69 | 15.44/0.73 | 20.84/0.40 | 20.39/0.66 | 20.10/0.65 |
| | K-SVD | 18.74/0.84 | 20.24/0.62 | 19.15/0.77 | 18.31/0.89 | 19.35/0.85 | 18.66/0.85 | 18.31/0.89 | 21.56/0.50 | 21.08/0.74 | 20.88/0.72 |
| | BM3D | 18.41/0.82 | 19.84/0.57 | 18.70/0.74 | 18.24/0.89 | 19.38/0.85 | 18.41/0.83 | 18.22/0.88 | 21.26/0.44 | 20.76/0.72 | 20.78/0.71 |
| 50 | Noisy | 14.17/0.72 | 14.16/0.48 | 14.14/0.64 | 14.17/0.80 | 14.16/0.70 | 14.18/0.73 | 14.16/0.79 | 14.16/0.36 | 14.15/0.51 | 14.09/0.51 |
| | NL-means | 15.16/0.52 | 18.43/0.41 | 16.21/0.44 | 14.11/0.62 | 16.72/0.69 | 15.24/0.58 | 14.15/0.61 | 19.94/0.32 | 18.70/0.50 | 18.74/0.55 |
| | K-SVD | 16.65/0.71 | 18.79/0.44 | 17.23/0.60 | 16.03/0.80 | 17.29/0.75 | 16.60/0.73 | 15.99/0.79 | 20.24/0.28 | 19.13/0.54 | 19.25/0.59 |
| | BM3D | 16.31/0.66 | 18.48/0.38 | 16.75/0.51 | 16.02/0.79 | 17.23/0.74 | 16.37/0.71 | 15.89/0.77 | 20.01/0.23 | 18.92/0.51 | 19.09/0.57 |
| 65 | Noisy | 11.90/0.61 | 11.87/0.36 | 11.90/0.51 | 11.90/0.71 | 11.88/0.59 | 11.86/0.62 | 11.91/0.70 | 11.89/0.26 | 11.87/0.39 | 11.87/0.40 |
| | NL-means | 14.41/0.43 | 17.86/0.36 | 15.58/0.37 | 13.20/0.52 | 15.66/0.61 | 14.45/0.50 | 13.25/0.51 | 19.20/0.27 | 17.68/0.40 | 17.79/0.48 |
| | K-SVD | 15.34/0.57 | 17.90/0.30 | 16.07/0.43 | 14.59/0.70 | 16.05/0.65 | 15.34/0.62 | 14.53/0.69 | 19.57/0.18 | 17.87/0.35 | 18.12/0.47 |
| | BM3D | 14.87/0.48 | 17.82/0.28 | 15.66/0.34 | 14.28/0.66 | 15.88/0.64 | 15.05/0.58 | 14.19/0.63 | 19.54/0.15 | 17.89/0.35 | 18.12/0.48 |
| 80 | Noisy | 10.06/0.51 | 10.11/0.27 | 10.08/0.41 | 10.04/0.61 | 10.12/0.50 | 10.07/0.52 | 10.12/0.60 | 10.02/0.19 | 10.04/0.30 | 10.04/0.31 |
| | NL-means | 13.88/0.37 | 17.34/0.33 | 15.14/0.32 | 12.60/0.45 | 14.94/0.55 | 13.97/0.45 | 12.64/0.43 | 18.50/0.24 | 16.99/0.34 | 17.10/0.43 |
| | K-SVD | 14.40/0.44 | 17.35/0.22 | 15.37/0.31 | 13.57/0.60 | 15.09/0.56 | 14.45/0.51 | 13.46/0.56 | 19.23/0.14 | 17.19/0.24 | 17.18/0.37 |
| | BM3D | 14.04/0.35 | 17.45/0.23 | 15.14/0.25 | 13.21/0.53 | 14.97/0.55 | 14.26/0.48 | 13.07/0.49 | 19.28/0.11 | 17.31/0.25 | 17.45/0.41 |
| 95 | Noisy | 08.54/0.42 | 08.57/0.21 | 08.62/0.33 | 08.55/0.53 | 08.58/0.41 | 08.55/0.44 | 08.60/0.52 | 08.60/0.14 | 08.55/0.23 | 08.56/0.25 |
| | NL-means | 13.48/0.32 | 16.81/0.30 | 14.75/0.29 | 12.13/0.39 | 14.38/0.50 | 13.57/0.41 | 12.26/0.39 | 17.86/0.21 | 16.40/0.30 | 16.46/0.39 |
| | K-SVD | 13.74/0.32 | 17.03/0.17 | 14.89/0.22 | 12.73/0.48 | 14.35/0.47 | 13.79/0.42 | 12.77/0.47 | 18.91/0.10 | 16.68/0.17 | 16.50/0.28 |
| | BM3D | 13.50/0.25 | 17.11/0.18 | 14.78/0.18 | 12.40/0.41 | 14.40/0.49 | 13.72/0.41 | 12.43/0.39 | 19.13/0.09 | 16.95/0.19 | 16.91/0.35 |
| 100 | Noisy | 07.29/0.35 | 07.31/0.16 | 07.30/0.27 | 07.34/0.46 | 07.31/0.35 | 07.35/0.37 | 07.29/0.45 | 07.32/0.11 | 07.27/0.19 | 07.31/0.20 |
| | NL-means | 13.14/0.30 | 16.30/0.28 | 14.35/0.27 | 11.85/0.36 | 13.95/0.47 | 13.23/0.38 | 11.92/0.36 | 17.18/0.19 | 15.81/0.26 | 15.87/0.35 |
| | K-SVD | 13.33/0.24 | 16.80/0.14 | 14.59/0.17 | 12.13/0.38 | 13.80/0.40 | 13.22/0.32 | 12.19/0.37 | 18.67/0.09 | 16.40/0.13 | 15.99/0.22 |
| | BM3D | 13.21/0.20 | 16.92/0.16 | 14.50/0.13 | 11.95/0.33 | 13.99/0.44 | 13.33/0.35 | 11.96/0.31 | 19.02/0.08 | 16.66/0.15 | 16.49/0.31 |

Table 10: Denoising using NL-means, K-SVD and BM3D algorithms when applied on different *64x64* images from *Synthetic Test Dataset* using multiple noise levels

| Noise | Algorithm | 1 | 2 | 3 | 4 | 5 | 6 | 7 | 8 | 9 | 10 |
|---|---|---|---|---|---|---|---|---|---|---|---|
| 5 | Noisy | 34.07/0.58 | 34.05/0.93 | 34.21/0.70 | 34.22/0.66 | 34.31/0.75 | 34.03/0.92 | 34.06/0.00 | 33.98/0.95 | 34.09/0.93 | 34.04/0.73 |
| | NL-means | 40.59/0.68 | 33.00/0.91 | 41.47/0.80 | 37.45/0.71 | 41.78/0.95 | 34.85/0.97 | 46.38/0.00 | 29.71/0.85 | 33.10/0.92 | 28.26/0.71 |
| | K-SVD | 42.98/0.71 | 35.79/0.95 | 44.15/0.82 | 41.84/0.73 | 44.06/0.97 | 39.46/0.98 | 47.66/0.00 | 34.58/0.95 | 35.92/0.95 | 35.62/0.75 |
| | BM3D | 47.75/0.77 | 35.74/0.95 | 49.40/0.87 | 42.19/0.81 | 48.30/0.99 | 41.92/0.99 | 63.54/0.59 | 34.62/0.96 | 35.84/0.95 | 36.20/0.89 |
| 20 | Noisy | 22.13/0.35 | 22.16/0.51 | 22.14/0.26 | 22.06/0.45 | 22.15/0.16 | 22.10/0.58 | 22.20/0.00 | 21.96/0.59 | 22.13/0.51 | 22.01/0.57 |
| | NL-means | 31.22/0.56 | 26.98/0.57 | 31.62/0.60 | 29.26/0.59 | 33.78/0.78 | 28.78/0.85 | 34.44/0.00 | 25.72/0.57 | 27.11/0.59 | 25.63/0.61 |
| | K-SVD | 34.58/0.63 | 27.46/0.64 | 32.95/0.66 | 33.16/0.65 | 33.70/0.83 | 31.83/0.92 | 40.50/0.00 | 26.26/0.66 | 27.37/0.65 | 26.53/0.64 |
| | BM3D | 37.91/0.69 | 27.27/0.59 | 37.93/0.81 | 33.71/0.69 | 38.21/0.93 | 33.40/0.93 | 48.76/0.18 | 25.99/0.61 | 27.53/0.63 | 26.87/0.75 |
| 35 | Noisy | 17.04/0.23 | 17.31/0.28 | 17.19/0.10 | 17.20/0.34 | 17.28/0.06 | 17.34/0.40 | 17.24/0.00 | 17.14/0.34 | 17.31/0.27 | 17.18/0.43 |
| | NL-means | 26.80/0.45 | 25.00/0.42 | 27.86/0.46 | 26.11/0.53 | 29.76/0.57 | 24.76/0.72 | 29.58/0.00 | 23.85/0.44 | 24.89/0.42 | 22.95/0.51 |
| | K-SVD | 29.51/0.52 | 24.78/0.37 | 30.16/0.54 | 29.13/0.58 | 33.74/0.85 | 26.12/0.80 | 36.15/0.00 | 23.85/0.44 | 24.85/0.37 | 23.40/0.53 |
| | BM3D | 32.72/0.61 | 25.21/0.36 | 31.70/0.63 | 30.84/0.65 | 35.50/0.86 | 29.19/0.89 | 44.92/0.03 | 24.05/0.42 | 25.06/0.37 | 23.19/0.53 |
| 50 | Noisy | 14.11/0.16 | 14.22/0.17 | 14.26/0.06 | 14.09/0.26 | 14.10/0.03 | 14.08/0.27 | 14.17/0.00 | 14.17/0.22 | 14.34/0.16 | 14.24/0.32 |
| | NL-means | 24.14/0.37 | 23.77/0.36 | 26.09/0.39 | 23.36/0.47 | 26.33/0.39 | 22.35/0.60 | 26.81/0.00 | 22.84/0.38 | 23.70/0.34 | 21.09/0.41 |
| | K-SVD | 26.10/0.44 | 24.00/0.27 | 29.84/0.53 | 25.81/0.53 | 31.52/0.82 | 23.13/0.68 | 34.95/0.00 | 23.02/0.33 | 23.97/0.27 | 21.56/0.41 |
| | BM3D | 30.47/0.56 | 24.56/0.30 | 31.61/0.65 | 26.27/0.55 | 32.61/0.83 | 25.36/0.79 | 56.37/0.29 | 23.35/0.32 | 24.51/0.31 | 20.88/0.36 |
| 65 | Noisy | 11.87/0.12 | 11.86/0.10 | 12.01/0.04 | 12.01/0.21 | 11.94/0.02 | 11.82/0.20 | 11.89/0.00 | 11.86/0.14 | 11.88/0.10 | 11.88/0.24 |
| | NL-means | 22.71/0.33 | 22.26/0.28 | 24.13/0.31 | 21.92/0.41 | 24.12/0.27 | 20.92/0.50 | 24.61/0.00 | 21.49/0.33 | 22.47/0.27 | 19.71/0.34 |
| | K-SVD | 24.83/0.39 | 23.67/0.23 | 29.46/0.54 | 25.16/0.49 | 29.90/0.75 | 21.60/0.58 | 33.27/0.00 | 22.05/0.27 | 23.74/0.23 | 20.25/0.34 |
| | BM3D | 27.74/0.49 | 24.14/0.26 | 29.80/0.58 | 24.79/0.51 | 30.89/0.78 | 23.35/0.70 | 44.77/0.18 | 22.46/0.29 | 23.94/0.25 | 20.29/0.35 |
| 80 | Noisy | 10.08/0.09 | 09.94/0.06 | 10.01/0.02 | 10.15/0.16 | 10.03/0.01 | 10.14/0.16 | 10.06/0.00 | 10.02/0.11 | 10.33/0.07 | 10.01/0.19 |
| | NL-means | 20.81/0.27 | 20.86/0.20 | 22.41/0.22 | 20.25/0.38 | 22.58/0.21 | 19.99/0.44 | 22.29/0.00 | 20.29/0.29 | 21.59/0.24 | 18.87/0.31 |
| | K-SVD | 23.04/0.33 | 22.92/0.20 | 28.59/0.50 | 22.80/0.45 | 29.27/0.69 | 21.08/0.55 | 30.15/0.00 | 22.11/0.24 | 23.58/0.20 | 19.37/0.28 |
| | BM3D | 26.32/0.45 | 23.40/0.20 | 28.08/0.49 | 22.62/0.47 | 29.61/0.72 | 22.33/0.65 | 52.38/0.16 | 22.31/0.28 | 23.97/0.24 | 19.32/0.27 |
| 95 | Noisy | 08.59/0.07 | 08.50/0.05 | 08.56/0.02 | 08.45/0.14 | 08.70/0.01 | 08.38/0.11 | 08.69/0.00 | 08.65/0.07 | 08.54/0.04 | 08.84/0.16 |
| | NL-means | 20.04/0.25 | 20.07/0.20 | 20.66/0.18 | 18.78/0.34 | 21.19/0.15 | 18.32/0.34 | 21.26/0.00 | 18.94/0.22 | 19.61/0.16 | 18.23/0.28 |
| | K-SVD | 22.15/0.29 | 23.21/0.23 | 27.04/0.48 | 22.14/0.43 | 28.53/0.72 | 19.29/0.42 | 29.21/0.00 | 21.16/0.19 | 22.59/0.18 | 18.90/0.23 |
| | BM3D | 26.12/0.44 | 23.37/0.24 | 26.60/0.44 | 21.83/0.46 | 27.47/0.62 | 20.49/0.52 | 45.76/0.10 | 21.13/0.20 | 22.99/0.22 | 19.31/0.26 |
| 100 | Noisy | 07.32/0.06 | 07.40/0.04 | 07.20/0.01 | 07.37/0.11 | 07.35/0.01 | 07.34/0.09 | 07.36/0.00 | 07.21/0.05 | 07.20/0.04 | 07.28/0.13 |
| | NL-means | 19.19/0.24 | 18.89/0.17 | 19.17/0.14 | 17.93/0.32 | 19.69/0.11 | 18.22/0.35 | 19.88/0.00 | 18.68/0.22 | 18.85/0.18 | 17.28/0.26 |
| | K-SVD | 21.66/0.26 | 22.67/0.19 | 26.02/0.45 | 20.39/0.38 | 27.60/0.68 | 19.63/0.47 | 28.53/0.00 | 21.22/0.19 | 22.46/0.19 | 18.77/0.24 |
| | BM3D | 24.63/0.41 | 22.70/0.20 | 25.00/0.39 | 20.96/0.43 | 26.70/0.57 | 20.54/0.53 | 38.72/0.17 | 21.37/0.20 | 22.92/0.20 | 19.09/0.26 |

Table 11: Denoising using NL-means, K-SVD and BM3D algorithms when applied on different *128x128* images from *Synthetic Test Dataset* using multiple noise levels

| Noise | Algorithm | 1 | 2 | 3 | 4 | 5 | 6 | 7 | 8 | 9 | 10 |
|---|---|---|---|---|---|---|---|---|---|---|---|
| 5 | Noisy | 34.20/0.30 | 34.12/0.93 | 34.18/0.41 | 34.17/0.39 | 34.19/0.61 | 34.13/0.73 | 34.10/0.00 | 34.25/0.99 | 34.13/0.93 | 34.08/0.66 |
| | NL-means | 42.66/0.35 | 35.73/0.95 | 43.30/0.60 | 40.64/0.44 | 38.57/0.71 | 38.17/0.76 | 46.54/0.00 | 26.45/0.92 | 35.79/0.95 | 25.14/0.65 |
| | K-SVD | 46.16/0.37 | 37.77/0.96 | 47.34/0.66 | 44.97/0.46 | 40.92/0.85 | 39.92/0.81 | 49.04/0.00 | 34.56/0.99 | 37.87/0.96 | 36.08/0.68 |
| | BM3D | 50.98/0.45 | 37.90/0.97 | 52.00/0.74 | 48.13/0.57 | 40.91/0.84 | 40.60/0.80 | 59.39/0.39 | 34.58/0.99 | 37.98/0.97 | 36.16/0.75 |
| 20 | Noisy | 22.03/0.19 | 22.17/0.57 | 22.06/0.13 | 22.16/0.25 | 22.08/0.09 | 22.07/0.37 | 22.09/0.00 | 22.11/0.83 | 22.15/0.56 | 22.18/0.62 |
| | NL-means | 32.32/0.29 | 28.30/0.74 | 33.13/0.37 | 31.44/0.35 | 32.91/0.41 | 30.78/0.58 | 34.36/0.00 | 23.83/0.80 | 28.12/0.72 | 24.34/0.64 |
| | K-SVD | 37.50/0.34 | 29.66/0.80 | 35.23/0.45 | 36.75/0.41 | 35.56/0.58 | 33.65/0.64 | 40.59/0.00 | 25.19/0.87 | 29.45/0.79 | 26.45/0.65 |
| | BM3D | 40.17/0.36 | 30.50/0.86 | 39.67/0.59 | 37.86/0.44 | 35.98/0.61 | 35.56/0.69 | 50.13/0.18 | 25.11/0.86 | 30.43/0.85 | 26.78/0.71 |
| 35 | Noisy | 17.30/0.14 | 17.30/0.33 | 17.22/0.05 | 17.22/0.18 | 17.25/0.03 | 17.23/0.24 | 17.28/0.00 | 17.23/0.63 | 17.28/0.33 | 17.27/0.57 |
| | NL-means | 28.13/0.25 | 24.69/0.49 | 29.11/0.27 | 27.48/0.31 | 29.36/0.24 | 26.60/0.47 | 29.82/0.00 | 21.28/0.65 | 24.82/0.51 | 22.16/0.62 |
| | K-SVD | 33.25/0.31 | 25.15/0.47 | 31.66/0.30 | 31.99/0.36 | 34.14/0.54 | 29.98/0.55 | 37.65/0.00 | 22.39/0.72 | 25.36/0.50 | 23.29/0.63 |
| | BM3D | 35.98/0.33 | 27.66/0.74 | 34.41/0.49 | 33.99/0.40 | 33.84/0.52 | 31.69/0.61 | 45.46/0.06 | 22.08/0.71 | 27.71/0.74 | 22.67/0.68 |
| 50 | Noisy | 14.21/0.10 | 14.19/0.20 | 14.11/0.03 | 14.12/0.14 | 14.10/0.01 | 14.17/0.17 | 14.06/0.00 | 14.19/0.49 | 14.11/0.20 | 14.21/0.52 |
| | NL-means | 25.41/0.21 | 23.02/0.37 | 26.33/0.20 | 24.74/0.27 | 26.37/0.14 | 24.10/0.39 | 26.66/0.00 | 19.62/0.53 | 23.01/0.38 | 20.27/0.59 |
| | K-SVD | 29.83/0.27 | 23.46/0.27 | 30.11/0.27 | 28.98/0.32 | 32.17/0.48 | 26.55/0.48 | 33.73/0.00 | 20.82/0.60 | 23.33/0.27 | 21.94/0.61 |
| | BM3D | 34.05/0.31 | 25.82/0.59 | 31.60/0.38 | 30.88/0.39 | 32.03/0.46 | 29.32/0.55 | 44.40/0.10 | 21.00/0.61 | 25.92/0.63 | 22.28/0.70 |
| 65 | Noisy | 11.92/0.08 | 11.79/0.13 | 11.82/0.02 | 11.85/0.11 | 11.88/0.01 | 11.82/0.13 | 11.84/0.00 | 11.82/0.36 | 11.94/0.13 | 11.84/0.45 |
| | NL-means | 23.33/0.18 | 21.60/0.27 | 24.10/0.15 | 22.81/0.24 | 24.22/0.09 | 22.15/0.33 | 24.18/0.00 | 18.18/0.39 | 21.73/0.28 | 18.20/0.54 |
| | K-SVD | 27.31/0.23 | 22.69/0.20 | 28.99/0.26 | 26.99/0.28 | 30.74/0.43 | 23.66/0.39 | 31.62/0.00 | 19.06/0.42 | 22.82/0.20 | 20.53/0.59 |
| | BM3D | 30.62/0.28 | 24.11/0.42 | 29.80/0.34 | 29.43/0.36 | 30.35/0.39 | 27.22/0.51 | 39.58/0.11 | 19.71/0.49 | 24.15/0.40 | 20.85/0.65 |
| 80 | Noisy | 10.06/0.06 | 10.01/0.09 | 10.13/0.01 | 10.06/0.09 | 10.09/0.01 | 10.12/0.10 | 10.05/0.00 | 10.11/0.27 | 10.05/0.09 | 10.12/0.40 |
| | NL-means | 21.59/0.16 | 20.51/0.23 | 22.46/0.12 | 21.30/0.21 | 22.68/0.06 | 20.73/0.28 | 22.76/0.00 | 17.31/0.31 | 20.56/0.22 | 16.41/0.47 |
| | K-SVD | 24.64/0.19 | 22.34/0.18 | 28.42/0.24 | 25.54/0.27 | 29.71/0.40 | 22.25/0.34 | 30.00/0.00 | 17.95/0.28 | 22.32/0.16 | 19.25/0.56 |
| | BM3D | 28.19/0.26 | 23.42/0.33 | 28.95/0.32 | 28.05/0.34 | 29.32/0.35 | 24.61/0.44 | 42.38/0.20 | 18.70/0.37 | 23.38/0.33 | 19.71/0.62 |
| 95 | Noisy | 08.57/0.05 | 08.64/0.07 | 08.51/0.01 | 08.59/0.07 | 08.62/0.00 | 08.63/0.07 | 08.54/0.00 | 08.58/0.22 | 08.59/0.07 | 08.60/0.35 |
| | NL-means | 20.30/0.14 | 19.45/0.19 | 20.84/0.09 | 20.05/0.19 | 20.99/0.04 | 19.65/0.24 | 21.16/0.00 | 16.54/0.26 | 19.55/0.19 | 15.29/0.41 |
| | K-SVD | 23.37/0.16 | 21.86/0.15 | 26.92/0.21 | 24.59/0.26 | 28.07/0.36 | 21.27/0.30 | 28.04/0.00 | 17.33/0.22 | 22.01/0.13 | 18.09/0.52 |
| | BM3D | 26.68/0.22 | 22.66/0.25 | 27.17/0.24 | 26.73/0.31 | 27.99/0.31 | 23.70/0.41 | 36.52/0.06 | 18.03/0.30 | 22.76/0.24 | 19.12/0.59 |
| 100 | Noisy | 07.37/0.04 | 07.33/0.05 | 07.39/0.01 | 07.32/0.06 | 07.35/0.00 | 07.26/0.06 | 07.30/0.00 | 07.23/0.17 | 07.26/0.05 | 07.34/0.30 |
| | NL-means | 19.22/0.13 | 18.55/0.15 | 20.04/0.08 | 18.92/0.18 | 19.86/0.03 | 18.69/0.21 | 19.74/0.00 | 15.96/0.23 | 18.66/0.17 | 14.38/0.35 |
| | K-SVD | 22.40/0.14 | 21.87/0.15 | 26.30/0.22 | 22.95/0.23 | 27.68/0.33 | 20.77/0.27 | 27.56/0.00 | 16.96/0.17 | 21.28/0.14 | 16.57/0.45 |
| | BM3D | 26.20/0.21 | 22.52/0.20 | 26.91/0.27 | 25.76/0.30 | 27.78/0.30 | 22.88/0.36 | 34.48/0.05 | 17.57/0.24 | 22.47/0.25 | 18.02/0.55 |

Table 12: Denoising using NL-means, K-SVD and BM3D algorithms when applied on different *256x256* images from *Synthetic Test Dataset* using multiple noise levels

| Noise | Algorithm | 1 | 2 | 3 | 4 | 5 | 6 | 7 | 8 | 9 | 10 |
|---|---|---|---|---|---|---|---|---|---|---|---|
| 5 | Noisy | 34.18/0.16 | 34.14/0.89 | 34.15/0.21 | 34.15/0.22 | 34.20/0.93 | 34.11/0.87 | 34.15/0.00 | 34.13/0.99 | 34.15/0.89 | 34.18/0.63 |
| | NL-means | 44.50/0.19 | 38.82/0.95 | 44.66/0.30 | 43.16/0.25 | 30.54/0.83 | 30.76/0.70 | 46.73/0.00 | 26.01/0.94 | 38.89/0.96 | 26.58/0.63 |
| | K-SVD | 48.08/0.20 | 40.07/0.96 | 48.62/0.34 | 47.24/0.27 | 35.14/0.94 | 35.82/0.88 | 50.15/0.00 | 34.42/0.99 | 40.19/0.96 | 36.50/0.64 |
| | BM3D | 53.81/0.27 | 40.45/0.96 | 53.52/0.46 | 51.99/0.36 | 35.45/0.94 | 36.79/0.90 | 61.62/0.44 | 34.48/0.99 | 40.54/0.97 | 37.24/0.76 |
| 20 | Noisy | 22.13/0.11 | 22.11/0.47 | 22.11/0.07 | 22.15/0.14 | 22.11/0.56 | 22.13/0.49 | 22.11/0.00 | 22.15/0.87 | 22.14/0.47 | 22.11/0.60 |
| | NL-means | 33.21/0.15 | 31.13/0.82 | 33.78/0.18 | 32.80/0.20 | 25.02/0.41 | 26.32/0.43 | 34.80/0.00 | 23.65/0.86 | 31.16/0.82 | 25.15/0.61 |
| | K-SVD | 38.96/0.18 | 32.08/0.83 | 37.69/0.24 | 38.44/0.23 | 26.35/0.56 | 27.79/0.50 | 41.63/0.00 | 24.91/0.91 | 31.95/0.82 | 26.44/0.62 |
| | BM3D | 42.05/0.19 | 33.27/0.88 | 42.01/0.33 | 41.74/0.26 | 25.85/0.47 | 27.89/0.48 | 49.53/0.16 | 25.14/0.91 | 33.28/0.88 | 27.54/0.69 |
| 35 | Noisy | 17.26/0.08 | 17.25/0.25 | 17.26/0.03 | 17.26/0.10 | 17.21/0.31 | 17.28/0.30 | 17.21/0.00 | 17.25/0.70 | 17.24/0.25 | 17.24/0.56 |
| | NL-means | 28.79/0.13 | 27.62/0.69 | 29.42/0.13 | 28.53/0.17 | 23.39/0.25 | 24.33/0.33 | 29.83/0.00 | 21.38/0.76 | 27.60/0.69 | 22.80/0.59 |
| | K-SVD | 35.18/0.17 | 26.71/0.56 | 33.40/0.15 | 34.28/0.21 | 24.00/0.23 | 25.58/0.33 | 36.87/0.00 | 22.23/0.81 | 26.87/0.56 | 23.31/0.60 |
| | BM3D | 37.93/0.17 | 30.22/0.81 | 36.04/0.25 | 37.62/0.25 | 23.83/0.19 | 25.84/0.33 | 43.44/0.05 | 22.29/0.82 | 30.28/0.81 | 23.98/0.62 |
| 50 | Noisy | 14.15/0.06 | 14.18/0.15 | 14.16/0.01 | 14.12/0.08 | 14.16/0.19 | 14.14/0.19 | 14.16/0.00 | 14.15/0.56 | 14.15/0.15 | 14.14/0.51 |
| | NL-means | 25.91/0.12 | 25.33/0.59 | 26.57/0.10 | 25.61/0.15 | 22.34/0.18 | 22.81/0.26 | 26.82/0.00 | 19.57/0.66 | 25.37/0.59 | 20.92/0.58 |
| | K-SVD | 32.12/0.15 | 24.35/0.34 | 31.71/0.13 | 30.88/0.18 | 23.23/0.11 | 24.52/0.26 | 34.23/0.00 | 20.58/0.70 | 24.48/0.35 | 21.65/0.59 |
| | BM3D | 35.43/0.16 | 28.43/0.74 | 33.48/0.22 | 34.55/0.22 | 23.24/0.09 | 25.06/0.27 | 44.87/0.10 | 21.11/0.75 | 28.43/0.74 | 22.91/0.64 |
| 65 | Noisy | 11.90/0.05 | 11.90/0.09 | 11.89/0.01 | 11.87/0.06 | 11.90/0.12 | 11.86/0.14 | 11.89/0.00 | 11.87/0.44 | 11.84/0.10 | 11.91/0.47 |
| | NL-means | 23.70/0.10 | 23.49/0.49 | 24.43/0.07 | 23.59/0.13 | 21.34/0.15 | 21.48/0.22 | 24.61/0.00 | 18.33/0.58 | 23.48/0.49 | 19.34/0.56 |
| | K-SVD | 29.23/0.13 | 23.39/0.25 | 30.37/0.12 | 28.67/0.17 | 22.88/0.08 | 23.31/0.21 | 32.48/0.00 | 19.28/0.59 | 23.53/0.26 | 20.55/0.58 |
| | BM3D | 33.23/0.15 | 26.69/0.64 | 31.99/0.21 | 31.80/0.21 | 23.01/0.06 | 24.58/0.25 | 41.47/0.10 | 20.31/0.70 | 26.94/0.65 | 21.85/0.63 |
| 80 | Noisy | 10.07/0.04 | 10.04/0.07 | 10.10/0.01 | 10.11/0.05 | 10.05/0.09 | 10.10/0.10 | 10.05/0.00 | 10.06/0.35 | 10.16/0.07 | 10.03/0.42 |
| | NL-means | 21.99/0.09 | 21.93/0.42 | 22.62/0.06 | 22.03/0.12 | 20.32/0.12 | 20.43/0.20 | 22.67/0.00 | 17.29/0.50 | 22.12/0.42 | 17.77/0.53 |
| | K-SVD | 26.68/0.11 | 22.93/0.21 | 29.06/0.12 | 27.16/0.15 | 22.57/0.07 | 22.08/0.17 | 29.95/0.00 | 17.87/0.46 | 22.96/0.20 | 19.61/0.56 |
| | BM3D | 31.33/0.14 | 25.55/0.54 | 30.76/0.16 | 30.64/0.20 | 22.87/0.05 | 24.09/0.23 | 38.61/0.08 | 19.67/0.66 | 25.54/0.53 | 20.94/0.60 |
| 95 | Noisy | 08.61/0.03 | 08.64/0.05 | 08.59/0.00 | 08.58/0.04 | 08.56/0.06 | 08.54/0.08 | 08.58/0.00 | 08.57/0.28 | 08.60/0.05 | 08.60/0.37 |
| | NL-means | 20.60/0.08 | 20.80/0.36 | 21.08/0.05 | 20.64/0.11 | 19.42/0.10 | 19.43/0.18 | 21.23/0.00 | 16.50/0.44 | 20.71/0.36 | 16.51/0.50 |
| | K-SVD | 25.15/0.10 | 22.55/0.18 | 27.86/0.11 | 25.83/0.14 | 22.25/0.05 | 21.38/0.15 | 28.73/0.00 | 16.78/0.34 | 22.57/0.19 | 18.90/0.55 |
| | BM3D | 29.76/0.13 | 24.68/0.44 | 30.29/0.16 | 29.29/0.18 | 22.79/0.05 | 23.65/0.22 | 38.67/0.10 | 19.17/0.62 | 24.67/0.46 | 20.31/0.60 |
| 100 | Noisy | 07.36/0.03 | 07.31/0.04 | 07.36/0.00 | 07.31/0.03 | 07.32/0.05 | 07.32/0.06 | 07.30/0.00 | 07.31/0.23 | 07.28/0.04 | 07.29/0.33 |
| | NL-means | 19.49/0.07 | 19.58/0.30 | 19.97/0.04 | 19.47/0.10 | 18.58/0.09 | 18.37/0.15 | 19.88/0.00 | 15.90/0.40 | 19.39/0.29 | 15.49/0.47 |
| | K-SVD | 23.78/0.08 | 22.18/0.17 | 27.09/0.10 | 24.79/0.13 | 22.08/0.05 | 20.57/0.13 | 27.42/0.00 | 15.99/0.25 | 22.04/0.16 | 17.83/0.53 |
| | BM3D | 28.74/0.12 | 23.83/0.36 | 29.72/0.14 | 28.45/0.16 | 22.78/0.05 | 22.83/0.20 | 37.21/0.08 | 18.68/0.59 | 23.56/0.35 | 19.82/0.61 |

## V. CONCLUSIONS AND FUTURE WORK

In this work, we have reached at possible conclusion in a positive way. The concerned results are now able to show image denoising facts and figures. In this section, we will justify the objectives of the research written in our proposal.

The followings are set of initial goals and objectives:
- To identify the different types of noisy signals over true signals.
- To study various image denoising algorithms.
- To find the image denoising processes.
- To simulate the MATLAB® for restoration processes.
- To make comparative analysis of various image denoising algorithms.

As shown in the previous section (Implementation), It is identified that noisy signal degrades image and disturbed true signal. To fulfil this objective, we have added Gaussian Noise to grayscale images. In the previous results section, tables showed noisy PSNR/SSIM of the image that has been degraded. As shown in the previous section (Results), we compared three image denoising algorithms. To meet this objective, we have had a detailed overview of these three denoising algorithms along with other recently proposed work in this domain. As shown in the previous section, we have developed a programming code that compare image restoration with respect to noisy, NL-means, K-SVD, and BM3D algorithms. In the section of implementation, we have tested over fifty images in three different sizes 64x64, 128x128, 256x256. All the images were also tested by adding noise using distinct noise levels. In this view, this objective has been fulfilled. In this research project, we have used MATLAB® for all processes including subjective and objective comparisons. This objective has been fulfilled and we can see the simulation results in section (implementation). As shown in the previous section results that different tables are given to show overall results provided by the three image denoising algorithms. To fulfil this objective, we have tested and compared three denoising techniques that are discussed in detailed in previous sections. The following features can be added in future to extend the functionality of this system:

- A comparative research can also be carried out to evaluate more denoising filtration.
- Since image denoising algorithm does not fully clean the image, this work can also be used to enhance the efficiency of such algorithms.
- Algorithm add or remove current parameters and constraints while denoising image.
- A new image denoising algorithm can be developed in which color images may also be denoised easily.